\documentclass[12pt]{article}
\usepackage{amsmath, amssymb, amsthm, thmtools}
\usepackage{graphicx,psfrag,epsf}
\usepackage{enumerate}
\usepackage{natbib}
\usepackage{url} 
\usepackage{algorithm2e}

\newcommand{\blind}{0}

\DeclareMathOperator*{\argmax}{arg\,max}

\renewcommand{\b}[1]{\mathbf{#1}}
\newcommand{\bs}[1]{\boldsymbol{#1}}
\newcommand{\s}[1]{\mathcal{#1}}
\renewcommand{\d}[1]{\mathbb{#1}}

\newcommand{\n}[1]{\mathrm{#1}}

\declaretheorem[name=Theorem]{theorem}
\declaretheorem[name=Proposition]{proposition}

\DeclareMathOperator*{\inprob}{\stackrel{P}{\longrightarrow}}
\DeclareMathOperator*{\inprobz}{\stackrel{P_0}{\longrightarrow}}

\DeclareMathOperator*{\indist}{\stackrel{d}{\longrightarrow}}

\newcommand{\GG}[1]{}

\begin{document}

\def\spacingset#1{\renewcommand{\baselinestretch}%
{#1}\small\normalsize} \spacingset{1}

\if0\blind{
  \title{\bf Beyond prediction: A framework for inference with variational approximations in mixture models}
  \author{T. Westling\thanks{
    The authors thank two referees and an associate editor for providing constructive feedback that helped them improve this manuscript. The authors also gratefully acknowledge grant 62389-CS-YIP from the United States Army Research Office, grants SES-1559778 and DMS-1737673 from the National Science Foundation, and grant number K01 HD078452 from the National Institute of Child Health and Human Development (NICHD).\hspace{.2cm}}\\
    Center for Causal Inference\\ University of Pennsylvania\\
 \and 
    T. H. McCormick \\
    Departments of Statistics \& Sociology \\ University of Washington}
  \maketitle
} \fi

\if1\blind
{
  \bigskip
  \bigskip
  \bigskip
  \begin{center}
    {\LARGE\bf Beyond prediction: A framework for inference with variational approximations in mixture models}
\end{center}
  \medskip
} \fi

\bigskip
\begin{abstract}
Variational inference is a popular method for estimating model parameters and conditional distributions in hierarchical and mixed models, which arise frequently in many settings in the health, social, and biological sciences. Variational inference in a frequentist  context works by approximating intractable conditional distributions with a tractable family and optimizing the resulting lower bound on the log-likelihood. The variational objective function is typically less computationally intensive to optimize than the true likelihood, enabling scientists to fit rich models even with extremely large datasets. Despite widespread use, little is known about the general theoretical properties of estimators arising from variational approximations to the log-likelihood, which hinders their use in inferential statistics. In this paper we connect such estimators to profile $M$-estimation, which enables us to provide regularity conditions for consistency and asymptotic normality of variational estimators. Our theory also motivates three methodological improvements to variational inference: estimation of the asymptotic model-robust covariance matrix, a one-step correction that improves estimator efficiency, and an empirical assessment of consistency. We evaluate the proposed results using simulation studies and data on marijuana use from the National Longitudinal Study of Youth.
\end{abstract}

\noindent%
\vfill

\newpage
\spacingset{1.45}


\section{Introduction}

Thanks to rapid improvements in data availability and user-friendly tools for data storage and manipulation, researchers from an ever-broader set of scientific disciplines now routinely analyze extremely complex, high-dimensional data. However, computing parameter estimates using stalwart statistical techniques such as maximum likelihood and Markov chain Monte Carlo can be a challenge in these settings and is often a bottleneck in practice.  In these situations, researchers often turn to computationally efficient approximations.
%
%

\emph{Variational approximations} are one method of approximating a likelihood function or posterior distribution that are increasingly popular across a range of scientific fields.   In public health, for example,~\citet{lee2016health} used a variational approximation to estimate a model for overall and hospital-specific trends in cesarean section rates.  In statistical genetics,~\citet{Raj573} used a variational approximation to a  multinomial model of allele frequencies across populations of individuals.~\citet{o2010discovering} used a variational approximation to a model of demographics and lexical choice in geo-tagged Twitter data.

Despite their popularity, variational approximations do not typically come with guarantees about the statistical properties of the resulting estimator. This drawback is particularly problematic when a scientist would like to interpret a parameter estimate, in which case estimator consistency is crucial, or report a confidence interval, in which case good coverage rates rely on the ability to accurately estimate the sampling distribution of the estimator.  
In this paper, we address the problem of inference using variational approximations.  We show that, in a wide range of  parametric mixture models, well-established theory from profile $M$-estimation provides an asymptotic lens through which we may understand the large-sample properties of parameter estimates resulting from variational approximations to the log-likelihood. Using the $M$-estimation framework, we derive conditions for consistency and asymptotic normality of variational estimators.

The theory we establish for variational estimators motivates us to also propose three methodological improvements to these estimators. First, we provide a consistent estimator of the asymptotic covariance matrix of variational estimators. Second, we introduce a one-step correction to the variational estimator that improves large-sample statistical efficiency. Third, we develop an empirical evaluation of estimator consistency for use when the theoretical calculations are intractable.  We demonstrate the importance of these methodological advances with two logistic mixture models of marijuana use by age among participants in the National Longitudinal Survey of Youth (NLSY). 



The remainder of the paper is organized as follows. This section  formally defines the class of models and variational estimators we study. Section~\ref{sec:thry} connects variational estimation to profile $M$-estimation and states our theoretical results. Section~\ref{sec:illustration} illustrates the general theoretical results in a few simple models. Section~\ref{sec:methods} presents our three methodological contributions. Section~\ref{sec:apply} evaluates our methods using simulated data and demonstrates an application to the NLSY.   Section~\ref{sec:concl} presents a discussion. Technical conditions and proofs of theorems are provide in the supplementary material. Code to replicate all of the empirical analyses in this paper are available at \url{https://github.com/tedwestling/variational_asymptotics}.

\subsection{Variational estimators}

In this paper, we consider inference for a Euclidean parameter $\theta$ in a parametric mixture model $p_{\theta}(x) = \int_{\mathcal{Z}} p_{\theta}(x , z)\, d\mu(z)$, where the marginal likelihood $p_{\theta}(x)$ is computationally expensive to compute.  Parametric mixture models have been used in a variety of scientific contexts. For example, mixed-membership models are a type of mixture model that have been used to model text~\citep{blei2003latent}, social networks~\citep{airoldi2008mixed}, population genetics~\citep{pritchard2000inference}, and scientific collaborations~\citep{erosheva2004mixed}.

Mixture models have been used in conjunction with both Bayesian and frequentist inferential frameworks.  In a frequentist setting, maximum likelihood (ML) estimation comes with guarantees of asymptotic efficiency and methods of conducting inference for many models.  These guarantees provide a degree of assurance for scientists that the point estimates and uncertainty intervals will behave in predictable ways.  

ML estimation can, however, be computationally burdensome.  When the integral in $p_{\theta}(x)$ must be approximated numerically, the cost of this computation increases exponentially with the dimension of the domain $\mathcal{Z}$ of the latent variable since  $p_{\theta}(x, z)$ needs to be evaluated at sufficiently many points to accurately approximate the integral.  This computational burden is a significant barrier for researchers who want to develop tailored mixture models to flexibly represent the dependencies in their data. As a result, a variety of approximate methods have been developed as alternatives to maximum likelihood.

Variational inference is an approximate method based on optimizing a lower bound for the original objective function. This lower bound is designed to eliminate the need for, or at least reduce the dimension of, any numerical integrals, thereby improving computational efficiency. Variational inference can be used in a frequentist context to approximate the log-likelihood or in a Bayesian context to approximate the posterior distribution. In this paper we focus on the former.  We will refer to estimators of $\theta$ resulting from optimizing a variational approximation to the log-likelihood as \emph{variational estimators}. 

Before providing formal definitions, we distinguish between two key aspects of the variational approximation.  First, we can evaluate the properties of the optimizer of the variational lower bound. Second, we could consider the tightness of the variational lower bound to the true objective function. These questions are related.  Demonstrating tightness of the lower bound is one way to control the difference between the true and variational optimizers, for example. However, a tight variational lower bound is not a necessary condition for good behavior of the variational estimator, and indeed does not hold in many settings where the variational estimator performs well. In this paper, we address the former of these two components, i.e.\ the properties of the optimizer of the variational lower bound.

We now move to a formal definition of variational estimation. \cite{blei2017review} presents a thorough introduction to variational inference and many relevant references. Let $X_1, \dotsc, X_n$ be observed $p$-variate data generated independently and identically from a distribution $P_0$ on a sample space $\s{X}$. Let $\mathcal{P} = \{P_{\theta}: \theta \in \Theta\}$ be a statistical model, where $\Theta$ is an open subset of $\mathbb{R}^d$ and each $P_{\theta}$ has a density $p_{\theta}(x) = \int_{\mathcal{Z}} p_{\theta}(x , z)\, d\mu(z)$. Here $\mu$ is a dominating measure on $\mathcal{Z} \subseteq \mathbb{R}^k$. We can conceptualize this data-generating process as first drawing independent \emph{latent} random variables $Z_1, \dotsc, Z_n$ from the marginal distribution $p_{\theta, Z}(z) = \int_{\s{X}} p_{\theta}(x,z) \, dx$, then drawing each $X_i$ given $Z_i$ from the conditional distribution $p_{\theta, X | Z}(x \mid Z_i) = p_{\theta}(x , Z_i) / p_{\theta, Z}(Z_i)$. Of these, we only observe $X_1, \dotsc, X_n$.

We are most interested in cases where $p_{\theta}(x)$ cannot be written in closed-form in terms of elementary functions, as in many generalized linear mixed models \citep{mcculloch2001generalized} and non-linear hierarchical models \citep{davidian1995nonlinear, goldstein2011multilevel}. In these cases, calculating the log-likelihood of the observed data, $\sum_{i=1}^n \log p_{\theta}(X_i)$, and its derivatives with respect to $\theta$ requires numerical integration. When the dimension of the latent variable is large, these numerical integrals are computationally expensive.

Variational inference parameter estimates are obtained by maximizing a criterion function motivated as follows. Denote by $\s{Q}_0$ the set of densities dominated by $\mu$, and by $\s{Q}_0^{\s{X}}$ the set of all conditional densities dominated by $\mu$ for all $x\in\s{X}$; that is, all $s : \s{Z} \times \s{X} \to \d{R}$ such that $s( \cdot \mid x) \in \s{Q}_0$ for all $x \in \s{X}$. Suppose that $P_0 \in \s{P}$, so that $P_0 = P_{\theta_0}$ for some $\theta_0 \in \Theta$. Then $\theta_0$ and the true conditional distribution of the latent variable $\pi_{\theta_0}(z \mid x) := p_{\theta_0}(x,z) / p_{\theta_0}(x)$ can be represented as 
\begin{equation}
(\theta_0, \pi_{\theta_0}) = \argmax_{\theta \in \Theta, s \in \s{Q}_0^{\s{X}}} E_{P_0} \left[ \int_{\s{Z}}\log \left(\frac{p_{\theta}(X, Z)}{s(Z \mid X)} \right) s(Z \mid X) \, d\mu(Z)\right].\label{eq:vem_conditional}
\end{equation}
To see this, first define
\[ f_0(\theta, s) := E_{P_0}\left[-D_{KL}(s(\cdot \mid X) \| \pi_{\theta}(\cdot \mid X)) \right] =  E_{P_0}\left[\int_{\s{Z}}\log\left(\frac{\pi_{\theta}(Z \mid X)}{s(Z \mid X)}\right)s(Z \mid X) \, d\mu(Z) \right] ,\]
where $D_{KL}$ denotes the Kullback-Leibler (KL) divergence. Thus, $f_0(\theta, s)$ is the expected KL divergence between $s(\cdot \mid X)$ and $\pi_{\theta}(\cdot \mid X)$.
By Gibbs' inequality, $f_0(\theta, s) \leq 0 = f_0(\theta_0, \pi_{\theta_0})$ for all $(\theta, s) \in \Theta \times \s{Q}_0^{\s{X}}$. Next, note that  $\theta_0$ maximizes $\theta \mapsto g_0(\theta) := E_{P_0}[ \log \frac{p_{\theta}(X)}{p_0(X)}] = -D_{KL}(p_0 \| p_{\theta})$. Therefore, $(\theta_{0}, \pi_{\theta_{0}})$ maximizes $(\theta, s) \mapsto f_0(\theta, s) + g_0(\theta)$ over $\Theta \times \s{Q}_0^{\s{X}}$, and after some rearranging, we can see that this is equivalent to the representation in~\eqref{eq:vem_conditional}.

The expectation-maximization (EM) algorithm can be motivated by \eqref{eq:vem_conditional} by replacing the unknown $P_0$ with the empirical distribution and alternating between optimization over $\theta$ and $s$. Using similar reasoning to that presented above, this amounts to alternating between computing $\theta_{(t)} := \argmax_{\theta \in \Theta} \sum_{i=1}^n\int_{\s{Z}}\left[ \log p_{\theta}(X_i, Z)\right] \pi_{\theta_{(t-1)}}(Z \mid X_i) \, d\mu(Z)$, where $\theta_{(t-1)}$ is the previous value of $\theta$, and computing $\pi_{\theta_{(t)}}(\cdot \mid X_i)$ for each observed $X_i$.  However, if the marginal likelihood $p_{\theta}(x)$ cannot be written in terms of elementary functions, then neither can $\pi_{\theta}$, and hence the EM algorithm requires numerical integration.

To construct a variational approximation to the log-likelihood, we replace the optimization over $\s{Q}_0^{\s{X}}$ in \eqref{eq:vem_conditional} with an optimization over $\s{Q}^{\s{X}}$, where $\mathcal{Q}$ is a smaller  \emph{variational family} of distributions, and as before $\s{Q}^{\s{X}}$ is the set of conditional distributions over $\s{X}$ such that $s(\cdot \mid x) \in \s{Q}$ for each $x \in \s{X}$. For example, $\mathcal{Q}$ could consist of all independent products over each dimension of $z$ (known as mean-field variational inference), all multivariate Gaussian distributions, or all independent Gaussian distributions. For simplicity, we will assume throughout that $\mathcal{Q}$ is indexed by a finite-dimensional Euclidean parameter $\psi \in \bs\Psi$, so that every $s \in \mathcal{Q}^{\s{X}}$ can be identified with a density $s(\cdot \mid x) = q(\cdot; \psi(x))$. We note that in some cases even when $\mathcal{Q}$ is a semiparametric family, it can be shown that the optimal $q$ lies in a parametric sub-family with a known form, so that our results can still be applied (see, e.g.\ Section 5.3 of \citealp{wainwright2008graphical}). For families where this does not apply, our theory could be extended to incorporate semiparametric $\mathcal{Q}$.

Let $\bs\Psi^n$ denote the $n$-fold Cartesian product $\bs\Psi \times \cdots \times \bs\Psi$. For $\bs\psi \in \bs\Psi^n$ and $i \in \{1, \dotsc, n\}$, we will denote $\psi_i \in \bs\Psi$ the $i$th element of $\bs\psi$. Given the observed data $X_1, \dotsc, X_n$, $\bs\Psi^n$ then parametrizes the set of variational conditional distributions over $X_1, \dotsc, X_n$, and each $\psi_i$ parametrizes the variational conditional distribution $s(\cdot \mid X_i) = q( \cdot; \psi_i)$. Given $\mathcal{Q}$ and $\bs\Psi$, the variational estimator of $\theta$, which we will denote $\hat\theta_{n}$, and the variational conditional estimators $\hat{\bs{\psi}}_n$ are the joint maximizers of the following objective function:
\begin{equation} (\hat{\theta}_n, \hat{\bs\psi}_n) := \argmax_{\theta \in \Theta, \bs\psi \in \Psi^n} \sum_{i=1}^n\int \log \left(\frac{p_{\theta}(X_i, Z_i)}{q(Z_i ; \psi_i)}\right) q(Z_i; \psi_i) \, d\mu(Z_i) = \argmax_{\theta \in \Theta, \bs\psi \in \Psi^n} \mathcal{L}_n(\theta, \bs\psi; \mathbf{X}_n).\label{VEM_estimate}\end{equation}
We note that we are implicitly assuming that the full variational distribution over $(Z_1, \dots, Z_n)$ factors as $\prod_{i=1}^n q(Z_i; \psi_i)$. However, since the true conditional distribution of $(Z_1, \dots, Z_n)$ given $(X_1, \dotsc, X_n)$ factors as $\prod_{i=1}^n \pi_{\theta_0}(Z_i \mid X_i)$, the optimal variational distribution will always factor as well, so this assumption comes with no loss of generality.

A crucial piece of motivation for our work is that, since $\mathcal{L}_n$ is typically not proportional to the log-likelihood, it is not clear what the asymptotic properties of the variational estimator $\hat\theta_n$ are.  In many circumstances, the variational estimator is used for prediction.  In such cases, scientists can evaluate the quality of the variational approximation using cross-validation or another held-out data technique.  If, however, a scientist would like to go beyond prediction and interpret the point estimator (or, critically, its uncertainty) produced by a variational approximation, not knowing the properties of the estimator is a substantial hindrance.  In particular, we would like to know whether $\hat\theta_n$ is consistent and, if it is consistent, what the asymptotic distribution of $\sqrt{n}(\hat\theta_n - \theta_0)$ is. 

Asymptotic properties of variational estimators have been studied in depth for certain specific models, yielding positive results regarding the consistency of variational  estimators for Gaussian mixture models \citep{wang2006normal}, exponential family models with missing values \citep{wang2004exponential}, Poisson mixed models as the cluster size and number of clusters both diverge \citep{hall2011theory, hall2011asymptotic}, Markovian models with missing values \citep{hall2002adequacy}, and stochastic block models for social networks \citep{bickel2013}. Of particular note are \cite{hall2011theory} and \cite{hall2011asymptotic}, who derive sharp asymptotics for Poisson regression with random cluster intercepts as both the number of clusters and observations per cluster diverge. Our work is distinct from these results in two ways. First, we provide results at a general level rather than for a specific model. Second, we focus on the asymptotic regime where the number of clusters is diverging, but the number of observations per cluster is stochastically bounded.

More recently, researchers have begun developing general theoretical results for variational estimators. For example, \cite{pati2018optimality} studied finite-sample risk bounds for mean-field variational Bayes estimators in a very general setting, and applied their results to derive the rate of convergence of variational Bayes estimators in Latent Dirichlet Allocation and Gaussian mixture models. \cite{wang2018frequentist} provided sufficient conditions for a Bernstein-von Mises result for the variational Bayes posterior distribution. We note that both of these recent works are distinct from our goals here, which are to study the asymptotic properties of frequentist variational estimators.

\section{Variational approximations and $M$-Estimation}
\label{sec:thry}

In this section, we demonstrate the connection between $M$-estimators and variational inference. The key for this connection is using a profile version of the variational objective function. Viewing variational inference in this way unlocks a deep and broad set of theoretical results developed for $M$-estimators.  We make this connection explicit in this section and then, in Section~\ref{sec:methods}, demonstrate how these theoretical results can be used to develop new methods for scientific practice.

\subsection{Variational estimation as $M$-Estimation}

We will study the general properties of the variational estimator $\hat\theta_n$ through the lens of $M$-estimation. An $M$-estimator of a parameter $\theta$ is the maximizer of a data-dependent objective function $M_n(\theta) = \frac{1}{n} \sum_{i=1}^n m(\theta; X_i).$ From \eqref{VEM_estimate} we can see that \[\mathcal{L}_n(\theta , \bs{\psi}_{n}; \b{X}_{n}) = \sum_{i=1}^n v(\theta, \psi_i; X_i) \mbox{ for } v(\theta, \psi; x) = \int \log \left( \frac{p_{\theta}(x , z)}{q(z; \psi)}\right)q(z; \psi) \, d\mu(z).\] 
Applying the theory of $M$-estimation to the vector $(\theta, \bs\psi_n)$ with $m(\cdot) = v(\cdot)$ is complicated due to the dependence on $\psi_i$, which are known as \emph{incidental} parameters specific to each data point.  The $\theta$, in contrast, are \emph{structural} parameters shared across all data \citep{Lancaster2000}. \citet{hall2011theory} dealt with this problem for Poisson mixed models by assuming the cluster size was growing with the number of observations, so that the incidental parameters effectively became structural. In our more general setting we could analogously assume that each observed data $X_i$ is composed of replicates $X_{i1}, \dotsc, X_{im}$ and let $m$ grow with $n$. However, this would limit the applicability of our results to only cases where clusters are very large.  Since, in practice, clusters are often small, we instead apply $M$-estimation to the \emph{profiled} variational objective.

In order to use the $M$-estimation framework for the variational estimator $\hat\theta_n$, we will express the optimization defined in \eqref{VEM_estimate} as a two-stage procedure, where first $\mathcal{L}_n$ is optimized with respect to $\bs{\psi}_{n}$ for each fixed $\theta$, then this \emph{profiled} function is optimized with respect to $\theta$. Furthermore, we note that optimizing $\mathcal{L}_n$ with respect to $\bs{\psi}_n$ for fixed $\theta$ is equivalent to optimizing each summand $v(\theta, \psi_i; x)$ with respect to $\psi_i$ for fixed $\theta$. Therefore, we will assume that for each $\theta \in \Theta$ and $P_0$-a.e.\ $x$, the map $\psi \mapsto v(\theta, \psi;x)$ possesses a unique point of maximum in $\bs\Psi$, which we will denote by $\hat\psi(\theta; x)$. We then define the profiled single-data objective function \[m(\theta; x) := \sup_{\psi \in \Psi} v(\theta, \psi; x) = v(\theta, \hat\psi(\theta, x); x).\]
Proposition~\ref{prop1} below asserts that, with this assumption, the variational estimator $\hat\theta_n$ of the model parameters from equation~\eqref{VEM_estimate} is equal to the maximizer of the profiled criterion function $\sum_{i=1}^n m(\theta; X_i)$. This result formally establishes the connection between variational inference and $M$-estimation that we will use throughout this article.
\begin{proposition}\label{prop1} Suppose that, for all $\theta \in \Theta$ and $P_0$-a.e.\ $x$, $\psi \mapsto v(\theta, \psi; X)$ possesses a unique maximizer in $\bs\Psi$, and that $\theta \mapsto \sum_{i=1}^n m(\theta; X_i)$ possesses at least one maximizer in $\Theta$. Then $\hat\theta_n\in\argmax_{\theta \in \Theta} \sum_{i=1}^n m(\theta; X_i)$.
\end{proposition}
The proofs of all results are provided in the supplementary material.

The representation of $\hat\theta_n$ provided by Proposition~\ref{prop1} now falls within the $M$-estimator framework.  We can therefore use the existing, well-studied asymptotic theory of $M$-estimators to better understand the asymptotic properties of variational estimators.  In the subsequent sections, we show that, using this representation, the theory for $M$-estimators yields general results for consistency and asymptotic normality for variational estimators.

\subsection{Consistency}

We first explore consistency using the $M$-estimator representation of the variational estimator. An important point that we will return to later is that, depending on the model and approximation, the estimator based on the variational lower bound may not be consistent for the truth. Hence, in what follows, we refer to $\bar\theta$ as the limit of $\hat\theta_n$, so that $\bar\theta = \theta_0$ if and only if $\hat\theta_n$ is consistent.

The population objective function $M_{0}(\theta) =  E_{P_0}[ m(\theta; X)]$ governs the asymptotic properties of the variational estimator $\hat\theta_n$. Under regularity conditions, $\hat\theta_n \inprobz \argmax_{\theta} M_{0}(\theta)$, so that if $M_{0}$ is uniquely maximized at $\theta_0$ then $\hat\theta_n$ is consistent for $\theta_0$, as we state below.
\begin{theorem}\label{vem_consistency}
Suppose the function $M_0(\theta) = E_{P_0}[ v(\theta, \hat\psi(\theta; X); X)]$ attains a finite global maximum at $\bar\theta$ and conditions (A1)-(A3) hold. Then $\hat\theta_n \inprobz \bar\theta$.
\end{theorem}
Regularity conditions (A1)-(A3) justify the application of Theorem 5.14 of \cite{van2000asymptotic} and are provided in the supplementary material. Condition (A1) requires that $v(\theta, \hat\psi(\theta; x); x)$ be upper semi-continuous in $\theta$ for a.e.\ $x$. This is implied, for instance, if $v$ is upper semi-continuous in $\theta$ and $\psi$ and $\hat\psi$ is continuous in $\theta$, for a.e.\ $x$. Condition (A2) requires that $v$ have a measurable and integrable local envelope function. Condition (A3) requires that $\hat\theta_n$ be contained in a compact with probability tending to one. If the parameter space is not compact, (A3) can often be established via a suitable compactification of the parameter space, as in \cite{van2000asymptotic} Example 5.16.

Here and throughout, we define $D_{\psi}$ and $D_{\theta}$ as the derivative operators with respect to $\psi$ and $\theta$, respectively. If $v$ and $\hat\psi$ are sufficiently smooth functions of $\theta$ for $P_0$-a.e.\ $x$ (see the supplementary material for additional details), then the Leibniz integral rule implies that $D_{\theta}M_0|_{\theta = \bar\theta} = E_{P_0}[D_{\theta} v |_{\theta = \bar\theta, \psi= \hat\psi(\bar\theta; X)}]$, and furthermore since $D_{\psi} v |_{\psi =\hat\psi(\theta; x)} = 0$ by definition of $\hat\psi$ as a maximizer,  $E_{P_0}[D_{\theta} v |_{\theta = \bar\theta, \psi= \hat\psi(\bar\theta; X)}] = 0$ as well. Therefore, a preliminary step in assessing whether $\hat\theta_n$ is consistent is to determine whether $E_{P_0}[D_{\theta} v |_{\theta = \bar\theta, \psi= \hat\psi(\bar\theta; x)}] = 0$. If it does not equal zero, then $\hat\theta_n$ cannot be consistent. If it does equal zero, and in addition $M_0(\theta)$ is strictly concave and regularity conditions (A1)--(A3) hold, then $\hat\theta_n$ is consistent.

In practice, it is often not possible to derive $\hat\psi(\theta; x)$ in closed form, which prevents a theoretical assessment of consistency of the variational estimator. This is the situation, for instance, in many generalized linear mixed models. In Section~\ref{sec:cons}, we propose an empirical method of assessing consistency that does not require explicit derivation of $\hat\psi(\theta; x)$.

\subsection{Asymptotic normality}

If the variational estimator $\hat\theta_n$ is consistent for $\bar\theta$ and additional regularity conditions  hold then $\sqrt{n}(\hat\theta_n - \bar\theta) \indist N(0, V(\bar\theta))$ where $V(\theta)$ is the sandwich covariance. Here and throughout, we denote by $D_{\bullet}^2$ the second derivative operator with respect to $\bullet$.
\begin{theorem}\label{vem_an}
Suppose $\hat\theta_n \inprobz \bar\theta$, a point of maximum of $M_0(\theta) = E_{P_0}[ m(\theta; X)]$, and conditions (B1)-(B4) hold. Then \[\sqrt{n}(\hat\theta_n - \bar\theta) \indist N_d(0, V(\bar\theta))\] where $V(\theta) = A(\theta)^{-1} B(\theta) A(\theta)^{-1}$ for
\begin{align} A(\theta) &=  E_{P_0}\left[D_{\theta}^2 m(\theta; X) \right] \label{eq:a_matrix}\\
B(\theta) &= E_{P_0}\left[ (D_{\theta} m(\theta; X)) (D_{\theta} m(\theta; X))^T\right]. \label{eq:b_matrix} \end{align}
\end{theorem}
In the next section we provide formulas for estimating the matrices $A$ and $B$ regardless of whether $m(\theta; X)$ is known explicitly.

Conditions (B1)--(B4), stated in the supplementary material, guarantee that $m(\theta; X)$ satisfies the conditions of \cite{van2000asymptotic} Theorem 5.23. Condition (B1) states that $\hat\psi(\theta; x)$ exists for all $\theta$ and a.e.\ $x$, and (B2) states that it is twice continuously differentiable in $\theta$ in a neighborhood of $\bar\theta$ for a.e.\ $x$. If for $\theta$ in a neighborhood of $\bar\theta$ and a.e.\ $x$, (i) $v$ is twice continuously differentiable in $\psi$, (ii) $D_{\psi}v|_{\theta, \hat\psi(\theta; x)} = 0$, (iii) $D_{\psi}^2 v$ is invertible, and (iv) $D_{\psi}v$ is twice continuously differentiable in $\theta$, then the implicit function theorem implies (B1) and (B2).

Condition (B3) requires that $v$ be twice continuously differentiable in $\theta$ in a neighborhood of $\bar\theta$ and $\hat\psi(\bar\theta; x)$ for a.e.\ $x$. The differentiability of $v$ required by this condition and the implicit function theorem from the previous paragraph depend on the smoothness of $p_{\theta}(x, z)$ and the variational density $q(z; \psi)$. For instance, by the Leibniz integral rule, if $p_{\theta}$ is twice continuously differentiable in $\theta$ at $x$ and for $q( \cdot; 
\psi)$-a.e.\ $z$, and its second derivative is dominated by a $q(\cdot; 
\psi)$-integrable function, then $v$ is twice continuously differentiable in $\theta$ at $\psi$ and $x$.

Finally, condition (B4) requires that $v$ and $\hat\psi$ be Lipschitz functions in neighborhoods of $\bar\theta$ and $\hat\psi(\bar\theta; x)$ for every $x$, and that their Lipschitz constant be bounded by a square-integrable function of $x$. The Lipschitz property of $v$ and $\hat\psi$ for fixed $x$ is implied by the  differentiability required by (B2) and (B3). Square-integrability of the Lipschitz constant as a function of $x$ is not guaranteed, but is a relatively mild requirement since the neighborhoods around $\bar\theta$ and $\hat\psi(\bar\theta; x)$ may be arbitrarily small.

\section{Illustrations of the general theory}\label{sec:illustration}

In this section, we illustrate the use of our theoretical results for assessing the consistency and asymptotic efficiency of variational estimators in two mixture models. For each model, we highlight the main features necessary to apply our general results, and leave detailed derivations for the supplementary material.

\subsection{Consistent and efficient variational estimation}

As our first illustration of our general theoretical results, we demonstrate that a variational estimator is consistent and efficient in an exponential mixture model. Suppose that each data unit $i$ consists of a vector of observations $X_i = (X_{i1}, \dotsc, X_{ip})$. Conditional on independent latent random variables $Z_1, \dotsc, Z_n$ each distributed as $\n{Exp}(\beta)$, these observations are generated independently as $X_{ij} \sim \n{Exp}(Z_i)$. The parameter vector is $\theta = \beta \in \d{R}^+$. This could serve, for instance, as a model of the lifetimes of clusters of memoryless units.

The marginal density of $X_i$ is $p_{\theta}(x)= \Gamma(p+1)\beta \left(\beta + \sum_{j=1}^p x_j\right)^{-(p+1)}$, and hence the true conditional distribution of $Z_i$ given $X_i$ is Gamma$(p + 1, \beta + \sum_{j=1}^p X_{ij})$. Therefore, any variational family of conditional distributions that includes the gamma family as a sub-class will yield a variational estimator $\hat\theta_n$ that is equal to the MLE. However, for the purpose of demonstrating our theoretical method of assessing consistency, it is illustrative to consider a variational class that does not include the true conditional distribution. We will show that in this example, using the mis-specified  variational class of log-normal distributions still yields a consistent, and even efficient, variational estimator.

Suppose the variational class is taken to be all log-normal distributions, parametrized by $\psi = (\mu, \sigma^2) \in \mathbb{R} \times \mathbb{R}^+ = \Psi$. Straightforward computation then gives
\[ v(\theta, \psi; x)  \propto \log\beta + (p+1)\mu - \left(\beta + \sum_{j=1}^d x_j\right) e^{\mu + \sigma^2 / 2}+ \log \sigma.\]
This is a smooth function, and by composition laws for concave functions, we can see that $v(\theta, \psi; x)$ is strictly concave in $\psi$ for each fixed $\theta$ and $x$ \citep{boyd2004convex}. Therefore, the unique zero of the gradient of $v$ with respect to $\psi$ is the unique $\psi$ maximizing $v$ for fixed $\theta$ and $x$. This gives $\hat\mu(\theta; x) = \log \frac{p+1}{\beta + \sum_{j=1}^p x_j} - (p+1)^{-1}/2$ and $\hat\sigma(\theta; x) = (p+1)^{-1/2}$. Thus, the profile objective function $m(\theta; x) = v(\theta, \hat\psi(\theta; x); x)$ can be written explicitly up to a constant as $\log \frac{\beta}{\left(\beta + \sum_{j=1}^p x_j\right)^{p+1}}$. Condition (A1) is satisfied because $m$ is smooth in $\theta$. Condition (A2) is satisfied because $\sup_{\theta} m(\theta; x) = c - p\log\left( \sum_{j=1}^p x_j\right)$ for some $c <\infty$, and the expectation of this expression is finite. Condition (A3), which requires tightness of $\hat\theta_n$, can be established either by restricting the parameter space to a compact, or by extending the parameter space to $[0, \infty]$ equipped with the metric $d(\beta_1, \beta_2) = |\arctan\beta_1 - \arctan\beta_2|$, as in \cite{van2000asymptotic} Example 5.16.

Since conditions (A1)--(A3) hold,   Theorem~\ref{vem_consistency} implies that $\hat\theta_n \inprob \bar\theta$, the point of maximum of $\theta\mapsto M_{0}(\theta) = E_{P_0}[m(\theta; X)]$. In this case, since $m(\theta; x)$ is equal up to a constant to the log-likelihood of a single observation, by a standard argument involving Jensen's inequality, $M_{0}(\theta)$ is uniquely maximized at $\theta_0$. Therefore, $\hat\theta_n$ is consistent even though the  variational class \emph{does not include} the true conditional distribution.

Conditions (B1)--(B3) are satisfied because both $v$ and $\hat\psi$ are smooth in $\theta$, and the second derivative of $m$ is bounded up to a constant in a neighborhood of $\theta_0$ by $(\sum_{j=1}^p x_j)^{-1}$, which is $P_0$-integrable. The Lipschitz condition (B4) is also satisfied because $v$ and $\hat\psi$ are differentiable with bounded derivatives in a neighborhood of $(\theta_0, \hat\psi)$ and $\theta_0$, respectively.  The asymptotic variance of $\sqrt{n}(\hat\theta_n - \theta_0)$, as implied by Theorem~\ref{vem_an}, is equal to $(1+2/p)\beta_0^2$.

\subsection{Inconsistent variational estimation}

We now consider an extension of the previous model in which a variational estimator is inconsistent. We keep an identical setup from the previous model, but now, we model the latent variable as $Z_i \sim \n{Gamma}(\alpha, \beta)$ rather than $\n{Exp}(\beta)$. This is a more flexible model indexed by the parameter $\theta = (\alpha, \beta) \in \d{R}^+ \times \d{R}^+$.

The marginal density of $X_i$ is now $p_{\theta}(x)= \Gamma(p+\alpha)\Gamma(\alpha)^{-1}\beta^{\alpha} \left(\beta + \sum_{j=1}^p x_j\right)^{-(p + \alpha)} $, and hence the true conditional distribution of $Z_i$ given $X_i$ is Gamma$(p + \alpha, \beta + \sum_{j=1}^p X_{ij})$. As before, for illustrative purposes we take the variational class to be all log-normal distributions, parametrized by $\psi = (\mu, \sigma^2) \in \mathbb{R} \times \mathbb{R}^+ = \Psi$. We now have
\[ v(\theta, \psi; x)  \propto \alpha\log\beta -\log\Gamma(\alpha)+ (p + \alpha)\mu - \left(\beta + \sum_{j=1}^d x_j\right) e^{\mu + \sigma^2 / 2}+ \log \sigma.\]
Once again, $v$ is a smooth function, and is strictly concave in $\psi$ for each fixed $\theta$ and $x$. Setting its derivative with respect to $\mu$ and $\sigma$ to zero and solving gives $\hat\mu(\theta; x) = \log \frac{p+\alpha}{\beta + \sum_{j=1}^p x_j} - (p+\alpha)^{-1}/2$ and $\hat\sigma(\theta) = (p+\alpha)^{-1/2}$. Thus, 
\begin{align*}
m(\theta;x) &\propto \alpha\log\beta -\log\Gamma(\alpha) + (p+\alpha) \log \frac{p+\alpha}{\beta + \sum_{j=1}^p x_j} - (p + \alpha) - \tfrac{1}{2}\log(p + \alpha) \\
&= \log p_{\theta}(x) - \log \Gamma(p + \alpha)- (\alpha + p) + (p + \alpha) \log(p + \alpha) -  \tfrac{1}{2} \log (p + \alpha).
\end{align*}
Conditions (A1)--(A3) can be checked for this example much as in the previous example. Therefore, Theorem~\ref{vem_consistency} again implies that $\hat\theta_n$ tends in probability to the point of maximum of $M_{0}(\theta)$. As before, $M_0$ is not available in closed form in terms of elementary functions. However, we have $M_0(\theta) = E_{P_0}[\log p_{\theta}(X)] + f(\alpha)$, where $f'(\alpha) > 0$ for all $\alpha$. Since $E_{P_0}[\log p_{\theta}(X)]$ is smooth and maximized at $\theta_0$, this implies that $D_{\theta}M_0|_{\theta=\theta_0} \neq 0$, so that $\theta_0$ cannot be the point of maximum of $M_0$. This shows that the variational estimator using a mis-specified log-normal conditional distribution is inconsistent in this example.

\begin{figure}[t]
\centering
\includegraphics[height=2in]{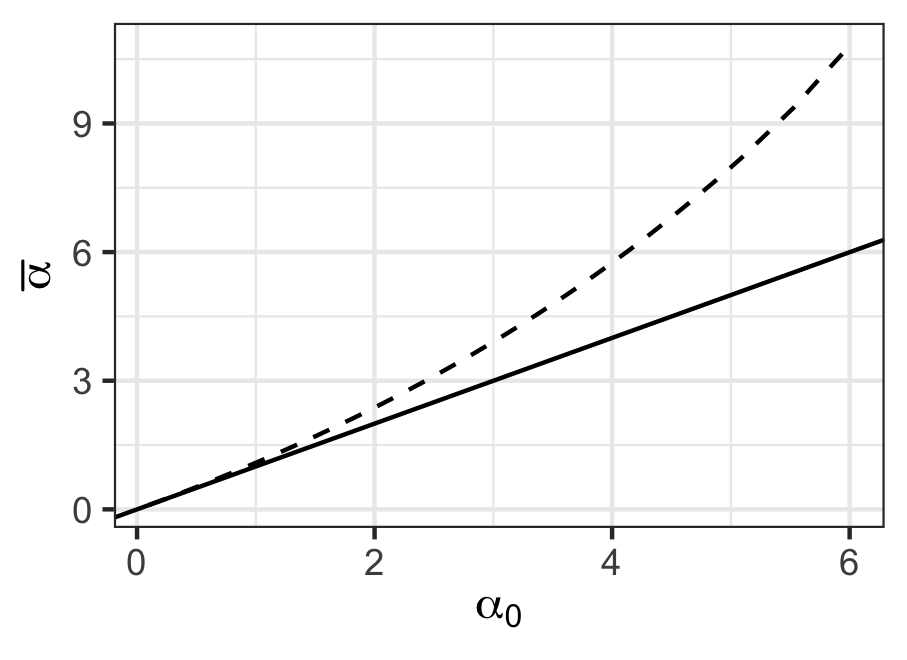}
\includegraphics[height=2in]{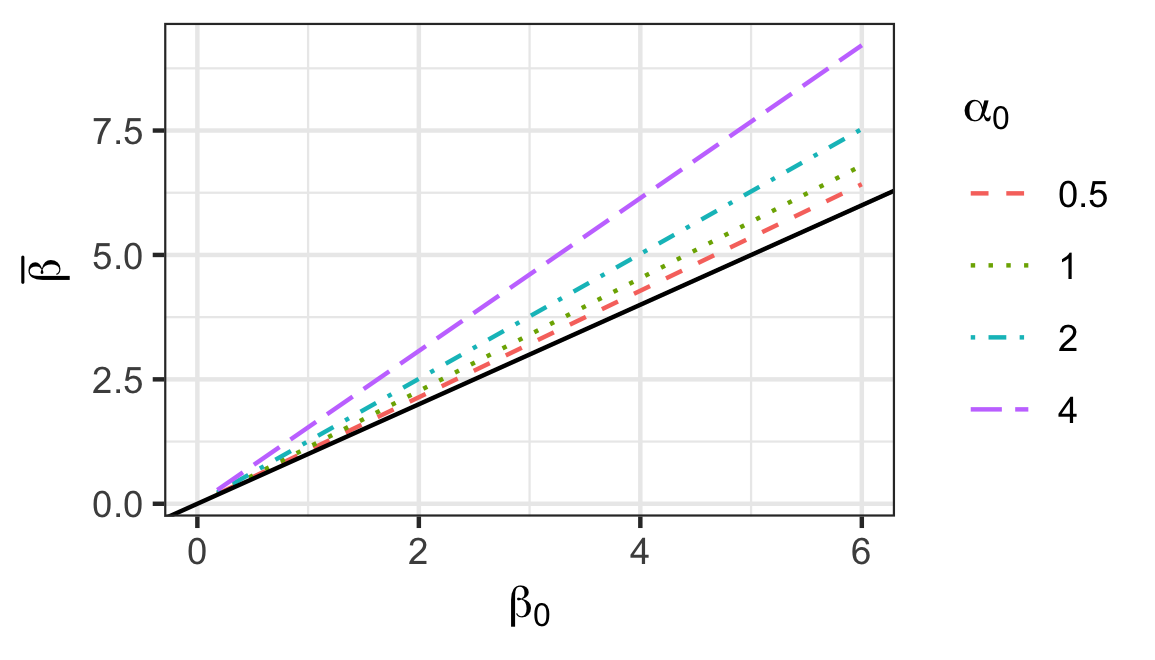}
\caption{Limits of the variational parameter estimates in the Exponential-Gamma mixture model using a mis-specified variational conditional distribution. The left display shows the limit $\bar\alpha$ as a function of the true $\alpha_0$ used to generate the data. Note that $\bar\alpha$ does not depend on $\beta_0$. The right display shows the limit $\bar\beta$ as a function of the true $\beta_0$ for four values of $\alpha_0$. The identity function is shown as a solid black line.}
\label{fig:biases}
\end{figure}

While the limit $\bar\theta$ is not available explicitly, we can approximate it using numerical integration and optimization. Figure~\ref{fig:biases} shows the limits of the variational estimators as a function of the true parameter value for $p=5$. The bias is small when $\alpha_0$ and $\beta_0$ are small, but increases as $\alpha_0$ and $\beta_0$ grow.

\section{Practical tools for inference with variational estimators}
\label{sec:methods}
We now propose three methodological innovations based on the asymptotic results from Section~\ref{sec:thry}.  First, we demonstrate how to leverage  asymptotic normality to enhance uncertainty estimators. Second, we show that a one-step correction can be applied to improve the efficiency of the variational estimator. Finally, we address the difficulty of theoretical assessment of consistency mentioned in Section~\ref{sec:thry}, providing a way to test the consistency of a variational estimator when theoretical calculations  are intractable.

\subsection{Sandwich covariance estimation}

We now discuss computation of consistent covariance estimators.  Recall that in practice, $m(\theta; X)$ is often not available in closed form.  Fortunately, the derivatives of $m(\theta; X)$ can be expressed in terms of the derivatives of $v(\theta, \psi; X)$, which are always available, because $v(\theta, \psi; X)$ is a result of the model and variational family used. Thus, using the chain rule, the asymptotic variance can be estimated whether or not $m(\theta; X)$ is available explicitly. We denote by $D_{\theta} v$ and $D_{\psi} v$ the first partial derivatives of $v$, and $D_{\theta\theta}^2 v$, $D_{\theta\psi}^2$, and $D_{\psi\psi}^2 v$ the second derivatives of $v$.

Concerning $D_{\theta} m (\theta; x)$, which appears in equation \eqref{eq:b_matrix}, since $\hat\psi(\theta; x)$ maximizes $v$ for fixed $\theta, x$, $D_{\theta} m(\theta; x) = D_{\theta} v(\theta, \psi; x) |_{\psi= \hat\psi(\theta; x)}.$  
For $D_{\theta}^2 m (\theta; x)$ in equation \eqref{eq:a_matrix},
\[D_{\theta}^2 m(\theta; x) = \left[D_{\theta\theta}^2 v  -  D_{\theta \psi}^2 v \left(D_{\psi\psi}^2 v \right)^{-1}  D_{\theta \psi}^2 v^T\right]_{\psi = \hat\psi(\theta; x)},\]
as we show in the supplementary material, where we abbreviate $v(\theta, \psi; X)$ as $v$ for presentation.
Replacing the appropriate derivatives in the definition of $V(\theta)$ with the above expressions and the population expectations with empirical ones gives a way to calculate the asymptotic covariance only knowing $v(\theta, \psi; X)$ and its derivatives (which can be calculated numerically), as opposed to $m(\theta; x)$, the computation of which involves optimization.

We now have $\hat{A}_n(\hat\theta_n)^{-1} \hat{B}_n(\hat\theta_n) \hat{A}_n(\hat\theta_n)^{-1} \inprob V(\bar\theta)$ where
\begin{align} \hat{A}_n(\theta) &= \frac{1}{n} \sum_{i=1}^n \left[D_{\theta\theta}^2 v  -  D_{\theta \psi}^2 v \left(D_{\psi\psi}^2 v\right)^{-1}  D_{\theta \psi}^2 v^T \right]_{\psi=\hat\psi_i, x= X_i},\label{eq:ahat} \\
 \hat{B}_n(\theta)&= \frac{1}{n} \sum_{i=1}^n \left[\left(D_{\theta} v \right) \left(D_{\theta} v\right)^T\right]_{\psi=\hat\psi_i, x=X_i}.\label{eq:bhat} \end{align}
Equations \eqref{eq:ahat} and \eqref{eq:bhat} provide a formula for constructing an asymptotic covariance matrix for the variational estimator $\hat\theta_n$. This covariance can be used to construct asymptotically calibrated Wald intervals, regions, and hypothesis tests about $\theta_0$ if $\bar\theta = \theta_0$. Furthermore, the sandwich covariance is model-robust in the sense that it is valid even if $P_0 \notin \s{P}$.

For an MLE under correct model specification, $A(\theta) = B(\theta)$ and the asymptotic covariance reduces to $A(\theta)^{-1}$, the inverse Fisher information matrix. In this case the sandwich covariance is only needed for model-robust uncertainty estimation. However, when $m$ is not proportional to the log-likelihood, as is often true with variational inference, $A$ and $B$ are not necessarily equal even under correct model specification. Therefore the sandwich covariance is necessary even if $P_0 \in \mathcal{P}$.

\subsection{One-step correction}

The variational estimator $\hat\theta_n$ is not guaranteed to be asymptotically efficient since the variational objective function need not be proportional to the log-likelihood. Hence while Wald-type intervals, regions, and tests using the sandwich estimator proposed in the last section will be asymptotically valid, they may be suboptimal since $\hat\theta_n$ may have larger asymptotic variance than the MLE. In these cases, a one-step correction to the variational estimator yields a more efficient estimator.

The one-step  estimator is
$\hat\theta_n^{(1)} = \hat\theta_n - I_n(\hat\theta_n)^{-1} S_n(\hat\theta_n)$, where $l(\theta; x) =  \log p_{\theta}(x)$ and
\begin{align*}
S_n(\theta) = \frac{1}{n}\sum_{i=1}^n D_{\theta}l(\theta; X_i), \hspace{2em} I_n(\theta) = \frac{1}{n}\sum_{i=1}^n (D_{\theta}l(\theta; X_i)) (D_{\theta}l(\theta; X_i))^T
\end{align*}
are the score and observed information at $\theta$. Under regularity conditions $\sqrt{n}(\hat\theta_n^{(1)} - \theta_0) \inprob N(0, I(\theta_0)^{-1})$ for $I(\theta_0)$ the Fisher information matrix, which is the same asymptotic distribution as the maximum likelihood estimator or posterior mean. 

Computing $S_n$ and $I_n$ require numerical integration in the same way that computing the  MLE would.  Indeed, the one-step correction is a single step of a Newton-Raphson algorithm for finding the MLE starting at $\hat\theta_n$. However, unlike finding the MLE, this one-step procedure only requires a single calculation of these quantities, so requires less computation than finding the exact MLE. Nevertheless, in some cases the one-step correction may not be computationally feasible for the same reasons that computing the MLE is not.

\subsection{An empirical test of the consistency of variational estimators}
\label{sec:cons}

In many cases, including generalized linear mixed models, neither $\hat\psi(\theta; x)$, $m(\theta; x)$, nor $M_0(\theta)$ are available analytically. This presents a challenge not present in the classical $M$-estimation scenario and seriously undermines the goal of theoretically evaluating the consistency of variational estimators.  Simulation studies could be used to assess consistency for any particular fixed, known truth, but would be computationally burdensome. 

Here, we propose a method for evaluating the consistency of a variational estimator at a single fixed parameter value $\theta^*$ when $m(\theta; x)$ is not available explicitly. Suppose that the data were generated from $P_0 = P_{\theta^*}$. Then a crucial condition for consistency of the variational estimator at $\theta^*$, as stated in Theorem~\ref{vem_consistency}, is that $M^*(\theta) := E_{\theta^*}[m(\theta; X)]$ be maximized at $\theta^*$. If $M^*$ is smooth and $\theta^*$ is in the interior of the parameter space, then $M^*$ being maximized at $\theta^*$ implies that $D_{\theta}M^*(\theta^*) = 0$. Furthermore, as long as $|D_{\theta} m(\theta; x)| \leq h(x)$ for all $\theta$ in a neighborhood of $\theta^*$ and $P_{\theta^*}$-a.e.\ $x$ for a $P_{\theta^*}$-integrable function $h$, then by the dominated convergence theorem, $D_{\theta} M^*(\theta^*) = E_{\theta^*}[ D_{\theta} m(\theta^*; X)] = E_{\theta^*}[D_{\theta}v(\theta^*, \hat\psi(\theta^*; X); X)]$. Our proposed method for numerically evaluating consistency of the variational estimator under $P_{\theta^*}$ is motivated by numerically assessing whether $E_{\theta^*}[D_{\theta}v(\theta^*, \hat\psi(\theta^*; X); X)] = 0$. Our method unfolds in the following steps.

\begin{enumerate}
    \item Fix $\theta^*$ and $b$ very large (for instance $10^4$ or $10^5$).
    \item For $j=1, \dotsc, b$:
    \begin{enumerate}[(a)]
    \item Simulate $X_j^* \sim P_{\theta^*}$.
    \item Find $\psi_j^* = \hat\psi(\theta^*; X_j^*)$ by numerically optimizing $\psi \mapsto v(\theta^*, \psi; X_j^*)$.
    \item Evaluate $G_j^* = D_{\theta} v |_{\theta^*, \psi_j^*, X_j^*}$.
    \end{enumerate}
    \item Test the null hypothesis that $E_{\theta^*}[G_j^*] = 0$ either using independent $t$-tests on each component or Hotelling's $T^2$ test on the entire vector.
\end{enumerate}
If the test rejects the null hypothesis then the variational estimator cannot be consistent; if not then one can be arbitrarily certain (with large enough $b$) that the mean score is zero at $\theta^*$. If a weakly significant $p$-value is found and it is unclear what to conclude, the experiment could be repeated with a larger $b$.

This method is a necessary, but not sufficient test of consistency. As we explain more below, asymptotically we expect our method to have few false negatives (indication that the estimator is inconsistent when it is actually consistent) but possibly false positives (indications that the estimator is consistent when it is actually inconsistent).  The first reason for potential false positives is that $E_{\theta^*}[G_j^*] = 0$ is a necessary but not sufficient condition for consistency. Even if its gradient is zero, $\theta^*$ it need not be a global maxima of the objective function. The second reason for potential false positives is that the method can only assess consistency at a single parameter value $\theta^*$ rather than on the entirety of the parameter space. Typically one will first use the variational algorithm to estimate $\hat\theta_n$, then use this method to  assess consistency at $\theta^* = \hat\theta_n$. If the estimator is consistent for every $\theta$ in a neighborhood of $\theta_0$ then for $n$ large enough $\hat\theta_n$ will be in that neighborhood and the method will not indicate inconsistency. On the other hand if $\hat\theta_n \inprobz \bar\theta \neq \theta_0$ then this method is approximately assessing whether the algorithm is consistent near $\bar\theta$. If the variational algorithm is consistent at $\bar\theta$ but not at $\theta_0$ then the method would indicate that the estimator is consistent when in fact it is not.  Despite the possibility of false positives, we do not know of any other practical ways to assess consistency of variational estimators when the limit objective is not available in closed form.

\section{Numerical studies}
\label{sec:apply}


In this section, we empirically evaluate the variational estimator, the sandwich covariance, and the one-step correction in mixed effects logistic regression models. In these models, theoretical assessment of the consistency and efficiency of variational inference is challenging because the profiled criterion function is not available in closed form. Hence, we turn to our empirical assessment of consistency and numerical studies to assess the properties of variational estimators.

We consider mixed effects logistic regression models -- first with random intercepts, then with random intercepts, slopes, and quadratic terms -- using data on marijuana use in adolescents in the United States from the National Longitudinal Survey of Youth 1997 \citep{nlsy97}.  The data consist of approximately yearly interviews of $n=8660$ youth from 1997 to 2012, with the number of interviews per youth ranging from four to sixteen. For youth $i$'s $j$th interview, we consider the binary outcome $Y_{ij}$ of whether the youth used marijuana in the thirty days preceding the interview. We focus on understanding the relationship between marijuana use, age, and sex. Since our goal is to understand the properties of variational estimators, we use the data, along with ``known'' parameter values, to simulate outcomes.  This way we can assess the accuracy of parameter estimates and coverage of uncertainty intervals.  We also use our methods to conduct an analysis of the real NLSY data.

The results indicate that variational estimators are not always consistent: in the first example the estimator is consistent for some parameters and not for others, and in the second example it is not consistent for any parameters. The first example also demonstrates that even when the variational estimator is consistent, it is not necessarily efficient. In either case the sandwich covariance matrix provides good confidence interval coverage rates and the one-step correction improves efficiency. The empirical evaluation of consistency correctly identifies inconsistency of the parameter vector as a whole, but not always inconsistency of individual parameters.

\subsection{Logistic regression with random intercepts}


First we consider logistic regression with random intercepts. Let $Z_i$ be a random intercept controlling each youth's overall propensity for marijuana use, $SEX_i$ be an indicator that the youth is male, and $AGE_{ij}$ be youth $i$'s age at interview $j$. Denote $p_{ij} = P(Y_{ij} = 1 \mid Z_i, SEX_{i}, AGE_{ij})$. Our first model for marijuana usage is then
\[ \mathrm{logit}(p_{ij}) = \begin{cases}  Z_i + \beta_0 + \beta_1 (AGE_{ij}/35) + \beta_2 (AGE_{ij}/35)^2,& SEX_i = 0 \\
Z_i + \beta_3 + \beta_4 (AGE_{ij}/35) + \beta_5 (AGE_{ij}/35)^2,& SEX_i = 1. \end{cases} \] 
Each youth's outcomes $Y_{i1}, \dotsc, Y_{in_i}$ are assumed conditionally independent given $Z_i$, and we model $Z_i$ as IID $N(0, \sigma^2)$. The parameter vector is $\theta = (\beta, \log(\sigma^2))$.
The inclusion of the quadratic effect of age is important because we expect that marijuana usage peaks some time in young adulthood and  decreases thereafter. This model form is similar to that used in the analysis of age-crime curves~\citep{fabio2011neighborhood}.

To estimate $\theta$ we consider a variational class of conditional distributions over $\b{Z}_{n}$ consisting of all independent Gaussian distributions. This is known as a Gaussian variational approximation (GVA). The variational parameters are $\psi_i = (m_i, \log s_i)$, $m_i$ being the mean and $s_i$ the standard deviation of the variational conditional distribution of $\gamma_i$. The variational objective involves one-dimensional numerical integrals. To optimize the variational objective function we use a variational EM algorithm using the statistical software \texttt{R} \citep{Rlang}. We used the \texttt{R} package \texttt{fastGHQuad} \citep{fastGHQuad} for numerical integration. In this case it is not possible to express the profiled objective function explicitly.

To evaluate our methods, we conducted a simulation study based on the NLSY data. For each of 1000 simulations, we draw a bootstrap sample of youth. Conditional on these youth's age and sex we simulated $Y_1, \dotsc, Y_n$ from the model, treating the variational estimate $\hat\theta_n$ for the data as the true parameter value $\theta_0$. We then estimated the model parameters and asymptotic covariance matrix using maximum likelihood with the \texttt{R} package \texttt{lme4} \citep{lme4}, the Gaussian variational approximation, and the one-step correction to the variational estimator. Finally, we used our proposed method to assess consistency of the variational algorithm at the estimated parameter value.
\begin{table}
\caption{Estimator variances in the logistic regression with random intercepts simulation.}\label{tab:random_intercept_variances}
\begin{center}
\begin{tabular}{rrrrrrrr}
  \hline
 & $\beta_0$ & $\beta_1$ & $\beta_2$ & $\beta_3$ & $\beta_4$ & $\beta_5$ & $\log(\sigma^2)$ \\ 
  \hline
MLE & 0.16 & 1.68 & 1.05 & 0.12 & 1.25 & 0.77 & $8.6\times 10^{-3}$ \\ 
 GVA & 0.23 & 1.90 & 1.19 & 0.19 & 1.46 & 0.91 & 0.61 \\ 
 One-step correction to GVA & 0.17 & 1.67 & 1.04 & 0.13 & 1.26 & 0.78 & 0.60 \\
   \hline
\end{tabular}
\end{center}
\end{table}

We first examine the accuracy of point estimates for regression fixed effects.  All three estimators concentrate on the true values of the fixed effects $\beta_0$ through $\beta_5$ (box plots are provided in the supplementary material). Table~\ref{tab:random_intercept_variances} shows the variance of the estimators for each of the seven model parameters. The variational estimator has slightly larger variance than the MLE, but the one-step correction nearly matches the variance of the MLE. Thus, as we asserted theoretically, the one-step correction is efficient as long as the variational estimator is consistent, even when the variational estimator is inefficient.

Moving now to the random intercept variance, the MLE concentrates on the true variance component, $\log(\sigma^2)$, while the variational estimator and one-step correction do not. Our empirical assessment of consistency described in Section~\ref{sec:cons} correctly identifies this inconsistency. The multivariate Hotelling test rejected in every simulation with $p < 10^{-16}$, correctly indicating that the population mean gradient of the entire parameter vector was significantly different from zero. Additionally, no more than 2.5\% of the marginal t-tests rejected at the 0.01 level for each of the fixed effects, in line with their apparent consistency, while every one of the 1000 simulations rejected the marginal $t$-test for the variance parameter with $p < 10^{-16}$. These results are better than what is guaranteed theoretically, since the theory does not guarantee that the marginal $t$-tests will accurately reflect the consistency or inconsistency of individual parameters.

We now move to examining uncertainty intervals. Table~\ref{tab:random_intercept_coverages} shows the estimated coverage of marginal 95\% Wald-type confidence intervals of the model parameters for each the three estimators. The coverage of the confidence intervals of the linear and quadratic age fixed effects using the sandwich covariance for the variational estimator and the inverse Fisher information for the one-step correction are within the Monte Carlo error of the nominal 95\%. The variational confidence intervals for the sex-specific intercepts $\beta_0$ and $\beta_3$ are too small at 90\%, likely because of the underestimation of $\sigma^2$, the variance of the random intercept. The coverage of the confidence intervals of $\log\sigma^2$ is close to zero for the variational estimator and one-step correction, which is not surprising given that the estimator is inconsistent. The coverage of $\log\sigma^2$ is not shown for the MLE because \texttt{lme4} does not provide an interval for this parameter.

\begin{table}
\caption{Coverage of 95\% confidence intervals in the logistic regression with random intercepts simulation.}\label{tab:random_intercept_coverages}
\begin{center}
\begin{tabular}{rrrrrrrr}
  \hline
 & $\beta_0$ & $\beta_1$ & $\beta_2$ & $\beta_3$ & $\beta_4$ & $\beta_5$ & $\log(\sigma^2)$ \\ 
  \hline
Maximum likelihood & 0.94 &  0.94 &  0.94 & 0.95 & 0.95 & 0.95 & -- \\ 
GVA + sandwich covariance & 0.90 &  0.94 & 0.94 & 0.90 &  0.94 & 0.94 & 0.09 \\ 
One-step correction to GVA &  0.93 & 0.94 & 0.94 & 0.91 & 0.95 & 0.95 & 0.02 \\ 
   \hline
\end{tabular}
\end{center}
\end{table}

\subsection{Logistic regression with random quadratics}

We now alter the model presented above to include random slopes and quadratic terms for each youth. 
\begin{figure}[t]
\centering
\includegraphics[height=3in]{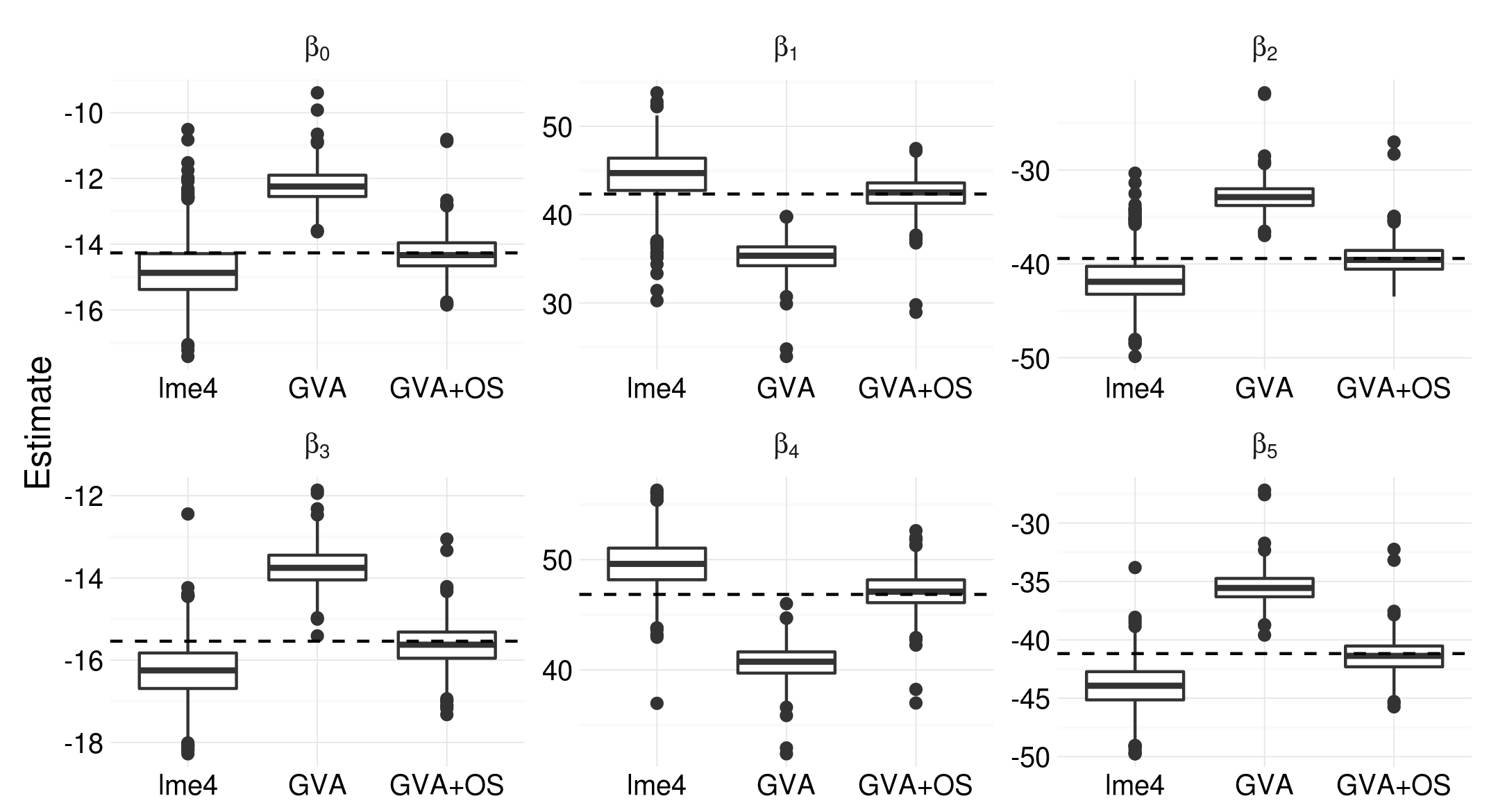}
\caption{Estimates of mean random effects from the logistic regression with random quadratics simulation study. The dotted line indicates the true parameter value. ``lme4" corresponds to estimate from the \texttt{lme4} package, which uses a Laplace approximation. ``GVA" stands for Gaussian variational approximation, and ``GVA+OS" refers to the one-step correction.}
\label{fig:rand_quad_sim_estimates}
\end{figure}
The random intercept model may not accurately capture the dependence structure of a single subject's marijuana use over time, since the random intercepts model implies an exchangeable marginal correlation structure, which is unrealistic given the longitudinal nature of the data. A more realistic model allows random slopes and quadratic terms as well, so that the latent variable $Z_i$ now has three components. For the conditional probability $p_{ij} = P(Y_{ij} =1| Z_i, SEX_i, AGE_{ij})$ we now have
\[ \n{logit}(p_{ij}) = \begin{cases} (Z_{i0} + \beta_0) + (Z_{i1} +\beta_1) (AGE_{ij}/35) + (Z_{i2} + \beta_2) (AGE_{ij}/35)^2,& SEX_i = 0 \\
(Z_{i0} + \beta_3) + (Z_{i1} +\beta_4) (AGE_{ij}/35) + (Z_{i2} + \beta_5) (AGE_{ij}/35)^2,& SEX_i = 1. \end{cases} \] 
Thus $\beta_0, \beta_1,$ and $\beta_2$ are the coefficients of the quadratic curve for the average female, and analogously for males. We model the random effects $\b{Z}_n$ as IID mean zero multivariate Gaussian with covariance matrix $\Sigma$. Once again we use MLE, a Gaussian variational approximation, and a one-step correction to the Gaussian variational approximation to estimate the average random effects and covariance matrix.


\begin{figure}[ht]
\begin{center}
\includegraphics[width=5in]{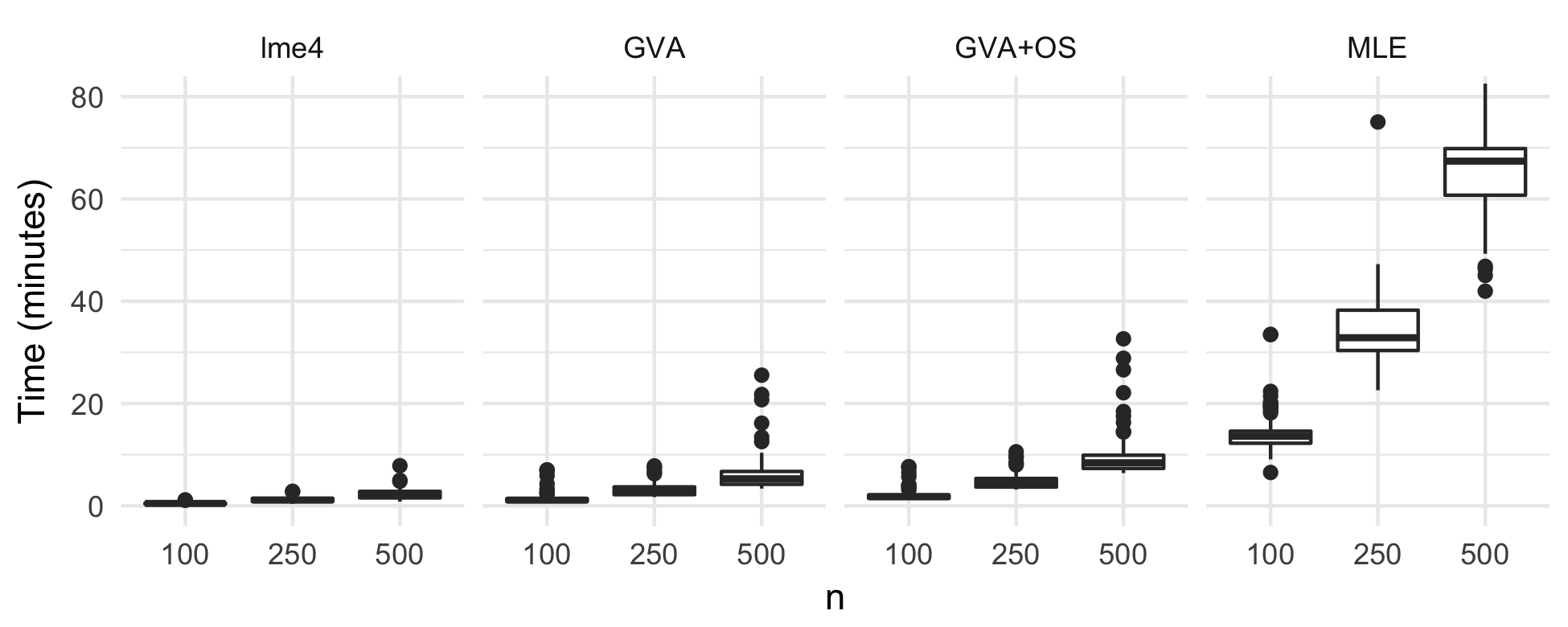}
\caption{Box plots of the computation time of four estimation methods of the logistic regression with random quadratics model. Three sample sizes are shown. ``lme4" refers to the \texttt{lme4} package, which uses a Laplace approximation to the marginal likelihood, ``GVA" stands for Gaussian variaitonal approximation, ``GVA+OS" refers to the one-step correction, and ``MLE" stands for maximum likelihood estimation, performed using L-BFGS-B optimization.}
\label{fig:timing_sims}
\end{center}
\end{figure}

The marginal likelihood for this model involves an intractable integral over $\d{R}^3$. GVA only requires numerical inegration over a one-dimensional integral, and is hence less computationally intensive than ML estimation. To compare the computational burden of these methods, we simulated 100 data sets each at sample sizes $n=100, 250$, and $500$, and used the \texttt{lme4} package, which uses a Laplace approximation to the log-likelihood, GVA, the one-step correction, and ML estimation to obtain estimates of the parameters in the logistic regression with random quadratics model. We used the implementation of the L-BFGS-B algorithm \citep{byrd1995bfgs} in the \texttt{optim} function in \texttt{R} \citep{Rlang} to compute the GVA and MLE. 

Figure~\ref{fig:timing_sims} shows box plots of the computation time in minutes of these four algorithms. The MLE was the most computationally expensive -- at sample size $n=500$, the average computation time was already 70 minutes. GVA and GVA+OS required an average of 6.3 and 9.7 minutes respectively to compute with $n=500$ observations. \texttt{lme4} was the most computationally efficient, requiring an average of 2.4 minutes.

We conducted a simulation study with the same structure as the study in the last section to compare the point estimates and CI coverage of \texttt{lme4} \citep{lme4}, GVA, and GVA+OS using all $8660$ observations. Box plots of the three estimators of the mean random effects are shown in Figure~\ref{fig:rand_quad_sim_estimates}. The pattern is very different from the random intercept model. The \texttt{lme4} estimates are slightly inconsistent (for random effects with dimension larger than one the \texttt{lme4} package uses a Laplace approximation to the likelihood). The GVA estimates are even more biased than the \texttt{lme4} estimates. Despite this, it appears that our proposed one-step corrected fixed effects are roughly centered around the true values. This is surprising since our theory does not guarantee that the one-step correction will be consistent when the variational estimate is not. All three estimators performed quite poorly in terms of estimating the covariance matrix of the random effects.

Table~\ref{tab:random_quad_coverages} shows the estimated coverage of 95\% CIs for the mean random effects for the three estimators. The variational sandwich coverage was close to 0 in every case due to the bias in the parameter estimate seen in Figure~\ref{fig:rand_quad_sim_estimates}. The \texttt{lme4} CIs also do not perform well, with substantially lower than desired coverage.  The one-step correction coverage is closest to the desired 95\%.  These intervals are conservative, containing the true value more than 95\% of the time.  

\begin{table}
\caption{Coverage of 95\% CIs in the logistic regression with random quadratics simulation.}\label{tab:random_quad_coverages}
\begin{center}
\begin{tabular}{rrrrrrrr}
  \hline
 & $\beta_0$ & $\beta_1$ & $\beta_2$ & $\beta_3$ & $\beta_4$ & $\beta_5$  \\ 
  \hline
Laplace approximation via \texttt{lme4} & 0.72 & 0.68 &  0.60  & 0.67  & 0.61 & 0.52 \\ 
GVA + sandwich & 0.01 & 0.01 & 0.00 & 0.013 & 0.01 & 0.00 \\ 
One-step correction to GVA  & 0.98 & 0.98 & 0.98 & 0.98 & 0.98 & 0.98 \\ 
   \hline
\end{tabular}
\end{center}
\end{table}

The multivariate Hotelling test of consistency soundly rejected for every simulation, correctly indicating that the variational parameter estimator is consistent.  In practice, therefore, while we would not be able to perform the same empirical evaluation as we have here (since we do not know the true model parameters to compute accuracy and coverage), we would have information that calls into question the viability of the GVA procedure for this model. The marginal $t$-tests of consistency for $\beta$  did not reject in the majority of simulations. Hence, while the marginal $t$-tests were an accurate diagnostic tool in the random intercept setting, they were not in the random quadratic setting.

\subsection{Analysis of marijuana use in NLSY97}

\begin{figure}
\includegraphics[width=6.5in]{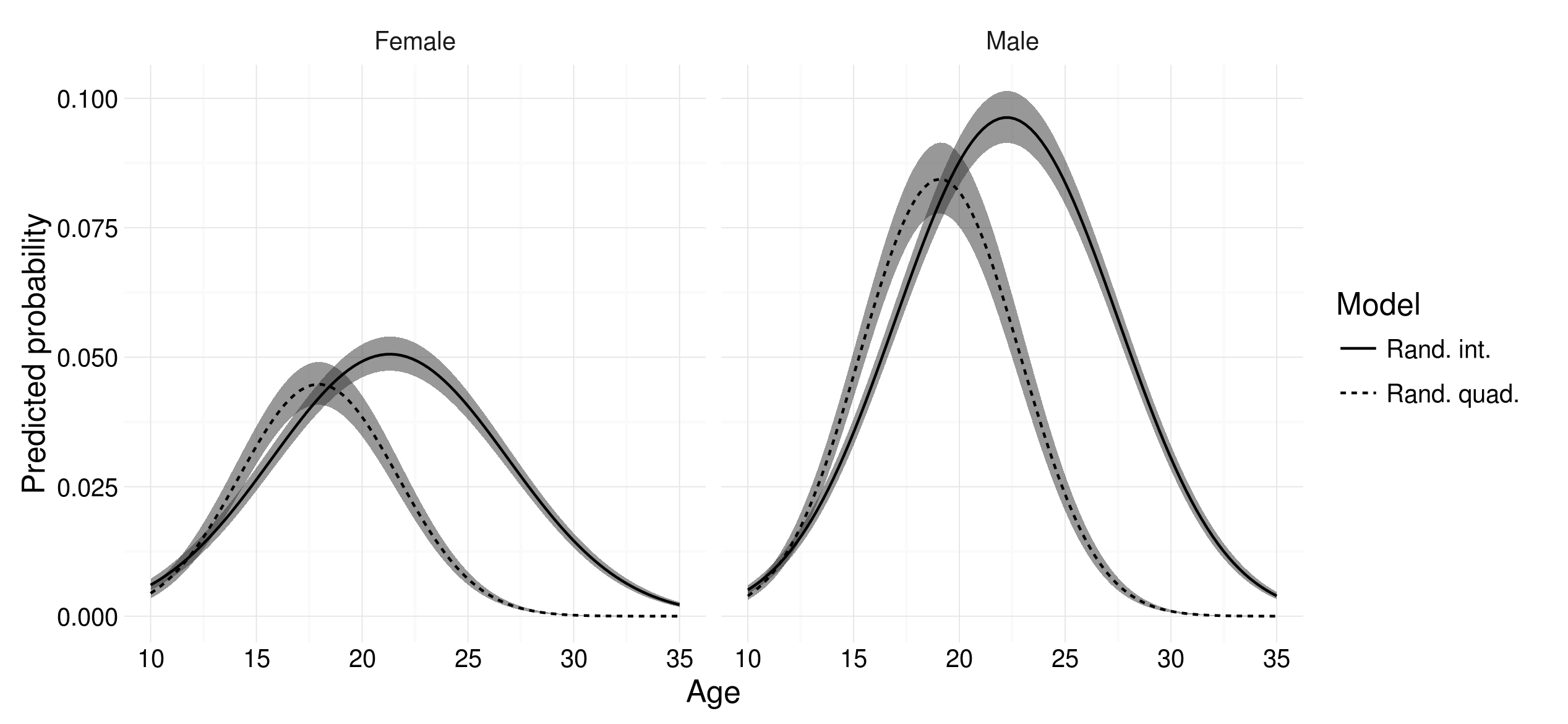}
\caption{One-step correction point estimates and pointwise 95\% confidence intervals of probability of having used marijuana in the past month. Curves on the left are for the average female, right are average male. Both the logistic regression with random intercepts and random quadratics are shown.}
\label{fig:both_nlsy97_pred}
\end{figure}

We used our one-step corrected estimator to assess the likelihood of marijuana usage by age and sex in the NLSY.  Figure \ref{fig:both_nlsy97_pred} shows the estimated mean curves as a function of age for both the random intercepts and random quadratics models and for both females and males. The average male and female from the random quadratics model have slightly faster increases, peak at younger ages, and decrease earlier than the average male and female from the random intercepts model. In both models the average male has higher overall probability and slightly later peak usage: in the random intercepts model, the estimated peak female usage probability occurs at 21.3 years (95\% CI: $[$18.1, 24.5$]$),  and peak male usage at 22.2 years (95\% CI: $[$19.9, 24.6$]$). In the random quadratics model, the estimated peak female usage probability occurs at 17.9 years (95\% CI: $[$15.9, 20.0$]$), and peak male usage at 19.1 years (95\% CI: $[$17.3, 20.8$]$).

\section{Discussion}
\label{sec:concl}
We have presented a general framework for understanding the properties of variational estimation for parametric mixture models.  The key insight of our work comes from representing the profiled variational objective function as an $M$-estimator.  Once we make this connection, we can leverage a rich toolkit of asymptotic and methodological results available for this context.  

The theory does not  guarantee that variational estimators are consistent, and it is often difficult to derive the profile objective function necessary to assess consistency. We proposed an empirical test of consistency based on estimating the gradient of the profile objective at a single parameter value. This proposed method worked well in practice, correctly indicating whether variational estimator is inconsistent in two generalized linear mixed models. 

We also used the asymptotic theory to propose a sandwich covariance estimator to provide calibrated confidence regions of variational estimators and a one-step correction to the variational estimator. Both of these methods work well when the variational estimator is consistent, and in fact the one-step correction exceeded our expectations by correcting some of the bias in fixed-effect variational parameter estimators in a logistic regression model with random quadratics.

Our theory is limited to models which are IID at some level. While this includes many hierarchical and longitudinal models, it excludes models for fully dependent time series, spatial data, and dyadic data. Extending the theory to cover those cases could be a fruitful next step. Additionally, we made the simplifying assumption that the variational class is of fixed and finite dimension, but our theory could be extended to other variational classes.

Our theory only provides results regarding the asymptotic behavior of the variational estimator $\hat\theta_n$ of the structural parameters $\theta_0$. It does not cover the behavior of the variational parameters $\hat\psi_i$, which govern the unit-specific variational conditional distributions of the latent variable $Z_i$ given the observed data $X_i$. The behavior of these parameters is important in settings where confidence regions with valid coverage are desired for the $Z_i$. This is the case, for instance, when the $Z_i$ correspond to specific fixtures of the real world, such as counties or schools.  However, we note that, since the variational family typically does not contain the true conditional distribution, it may be difficult to provide regions with good coverage of $Z_i$ using variational inference. By contrast, as we have demonstrated, $\hat\theta_n$ may be consistent even when the variational family does not contain the true conditional. This was one reason we chose to focus on the asymptotic behavior of $\hat\theta_n$. Nevertheless, we would conjecture that $\hat\psi_i$ given $X_i$ converges in distribution to 
\[ \argmax_{\psi \in \bs\Psi} \int \left( \log \frac{p_{\bar\theta}(X_i, z)}{q(z; \psi)}\right)q(z; \psi) \, dz \]
for each $i$, where $\bar\theta$ is the limit in probability of the variational estimator $\hat\theta_n$. We leave further discussion along these lines to future work.

We developed our theory in the context of a fixed variational family of conditional distributions. A natural question is whether our theory provides insight about which variational families yield consistency. Unfortunately, it appears to be difficult to address this question in a general manner. As a simple example, it would intuitively seem that if a given class of variational distributions yields a consistent estimator, then any enlargement of the class should also yield a consistent estimator. However, it is not clear whether this is true based on our theory. This would be an important topic of future research.

\section*{Supplementary material}
\label{SM}

Supplementary material includes proofs of Theorems 1 and 2, an intuitive explanation of the over-concentration of the variational Bayes posterior distribution, derivations related to the exponential mixture model, and additional simulation results.

\bibliographystyle{agsm}
\bibliography{paper-ref}

\clearpage
\begin{appendix}

\begin{center}\Huge{\textsc{Supplementary Material}}\end{center}
\section{Example of underestimated uncertainty for variational approximations}

As discussed in the main text, even when the variational Bayes posterior is consistent, it frequently underestimates the true posterior variance. This phenomenon can be explained intuitively using the KL divergence between multivariate normal distributions. Under regularity conditions, the posterior distribution of model parameters $\Pi_n(\theta | X_{1:n})$ looks approximately $N_D(\theta_0, \Sigma)$ as $n$ grows (where $\Sigma$ implicitly depends on $n$ because $\theta$ has not been appropriately rescaled). Often the variational distribution is asymptotically normal as well. However, if the variational class of distributions over model parameters only includes factored distributions, then the variational distribution can only be approaching independent normal distributions. The KL divergence between a normal distribution with mean $\mu$ and diagonal covariance matrix with $k$th diagonal entry $\sigma_k^2$ and a general multivariate normal is minimized when $\mu = \theta_0$ and $\sigma_k^2 = 1/(\Sigma^{-1})_{kk}$. However, using Schur complements we can see that  
\[\sigma_k^2 = \tfrac{1}{(\Sigma^{-1})_{kk}} = \Sigma_{kk} - \Sigma_{k\cdot} (\Sigma_{-kk})^{-1} \Sigma_{k\cdot}^T,\]
where $\Sigma_{k\cdot}$ is the $k$th row of $\Sigma$ omitting $\Sigma_{kk}$ and $\Sigma_{-kk}$ is the minor of $\Sigma$ removing the $k$th row and $k$th column. Assuming $\Sigma$ is positive definite, $\Sigma_{-kk}^{-1}$ is positive definite as well and hence $\Sigma_{k\cdot} \Sigma_{-kk}^{-1} \Sigma_{k\cdot}^T \geq 0$ with equality if and only if $\Sigma_{k\cdot} = \boldsymbol{0}$. Hence $\sigma_k^2$, the marginal variational posterior variance of $\theta_k$, is $\leq \Sigma_{kk}$, the true marginal posterior variance, with equality if and only if $\theta_k$ is not correlated in the posterior with any of the other model parameters. Thus, we should expect the variational Bayes posterior to underestimate the marginal uncertainty of any model parameter or latent variable that is asymptotically correlated with other model parameters or latent variable.  

\section{Proof of theorems}

Recall that $P_0$ is the true distribution, $\s{Q}$ is the variational family of distributions over the latent variable $Z$, which is parametrized by $\psi \in \Psi$, and 
\[v(\theta, \psi; x) = E_{\psi}\left[\log \frac{p_{\theta}(x, Z)}{q(Z; \psi)}\right] \]
is one term in the variational criterion function.

We first prove Proposition 1.
\begin{proof}[Proof of Proposition 1]
Suppose that $\hat\theta_n$ is not a maximizer of $\sum_{i=1}^n m(\theta; X_i)$. Let $\tilde\theta_n \in \argmax_{\theta \in \Theta} \sum_{i=1}^n m(\theta; X_i)$, which exists by assumption. Also by assumption, we may define $\hat\psi(\theta; x) := \argmax_{\psi \in \bs\Psi} v(\theta, \psi; x)$ for each $\theta \in \Theta$ and $x \in \s{X}$. Then by the definition of $m$,
\[\sum_{i=1}^n m(\tilde\theta_n; X_i) = \sum_{i=1}^n v(\tilde\theta_n,\hat\psi(\tilde\theta_n; X_i); X_i) >\sum_{i=1}^n m(\hat\theta_n; X_i) = \sum_{i=1}^n v(\hat\theta_n,\hat\psi(\hat\theta_n; X_i); X_i).\]
Now for each $i$, by definition of $\hat\psi(\hat\theta_n; X_i)$, $v(\hat\theta_n,\hat\psi(\hat\theta_n; X_i); X_i) \geq v(\hat\theta_n,\hat\psi_i; X_i)$, where $\hat\psi_i$ is the variational maximizer from equation (2) of the main text. Hence, $\tilde\theta_n \neq \hat\theta_n$ implies that
\[ \sum_{i=1}^n v(\tilde\theta_n,\hat\psi(\tilde\theta_n; X_i); X_i) > \sum_{i=1}^n v(\hat\theta_n,\hat\psi_i; X_i).\]
This is a contradiction, since $(\hat\theta_n, \hat{\bs\psi}_n)$ are defined as the joint maximizers of $\sum_{i=1}^n v(\theta,\psi_i; X_i)$.
\end{proof}

We now provide the list of assumptions we will need for consistency:
\begin{description}
\item[(A1)] The map $\theta \mapsto \hat\psi(\theta; x)$ is upper-semicontinuous a.s.-$P_0$.
\item[(A2)] There exists a $d > 0$  such that for all $\delta < d$ and $\eta \in \Theta$ the map
\[ x \mapsto \sup_{\substack{\theta \in B_{\delta}(\eta) \\ \psi \in \Psi}} v(\theta, \psi; x) \]
is measurable and
\[ E_{P_0} \sup_{\substack{\theta \in B_{\delta}(\eta) \\ \psi \in \Psi}} v(\theta, \psi; X)  < \infty.\]
\item[(A3)] There exists a compact set $K \subset \Theta$ such that $P_0(\hat\theta_n \in K) \to 1$.
\end{description}
We can now demonstrate Theorem 1.

\begin{proof}[Proof of Theorem 1]
Defining $m(\theta;x) = \sup_{\psi \in \Psi} v(\theta, \psi; x)$, the requirements of Theorem 5.14 of \citet{van2000asymptotic} are satisfied. Hence for all $\epsilon > 0$, $P_0(\|\hat\theta_n - \bar\theta\| \geq \epsilon \cap \hat\theta_n \in K) \to 0.$
Since $P_0(\hat\theta_n \in K) \to 1$ by assumption,
\begin{align*} P_0(\|\hat\theta_n - \bar\theta\| \geq \epsilon) &\leq P_0(\|\hat\theta_n - \bar\theta\| \geq \epsilon \cap \hat\theta_n \in K) + P_0(\|\hat\theta_n - \bar\theta\| \geq \epsilon \cap \hat\theta_n \in K^c) \\
&\leq P_0(\|\hat\theta_n - \bar\theta\| \geq \epsilon \cap \hat\theta_n \in K)  + P_0( \hat\theta_n \in K^c) \to 0.
\end{align*}
\end{proof}

We next lay out sufficient conditions for asymptotic normality. Throughout we will assume that for all $\theta$, $(\psi, x) \mapsto v(\theta, \psi; x)$ is a measurable function on the product measure space $\Psi \times \s{X}$, where $\Psi$ is equipped with Borel measure.
\begin{description}
\item[(B1)] For all $\theta$ and $P_0$-a.e.\ $x$, $v(\theta, \psi; x)$ is uniquely maximized at $\hat\psi(\theta; x)$ which is an element of $\Psi$, an open subset of $\b{R}^d$.
\item[(B2)] $\hat\psi$ is a measurable function of $x$ for all $\theta$ and twice continuously differentiable in a neighborhood of $\bar\theta$ for $P_0$-a.e.\ $x$.
\item[(B3)] $v$ is twice continuously differentiable in a neighborhood of $\bar\theta$ and $\hat\psi(\bar\theta; x)$  for $P_0$-a.e.\ $x$, and there exists a $P_0$-integrable function $\kappa$ such that for all $\theta$ in a neighborhood of $\bar\theta$ and $P_0$-a.e.\ $x$,
\[\left| \left( D_{\theta\theta}^2 v- (D_{\theta\psi}^2 v) (D_{\psi\psi}^2 v)^{-1} (D_{\theta\psi}^2 v)^T\right) (\theta, \hat\psi(\theta; x); x) \right| \leq \kappa(x).\] 
\item[(B4)] There exist $r > 0$, $s(x) >0$, $b_1(x)$ and $b_2(x)$ such that
\begin{enumerate}[(a)]
\item For all $x\in\s{X}$ and $\theta \in \s{B}_r(\bar\theta)$, $\hat\psi(\theta; x) \in \s{B}_{s(x)}(\hat\psi(\bar\theta; x))$
\item For all $x\in\s{X}$, $\theta_1, \theta_2 \in \s{B}_r(\bar\theta)$ and $\psi_1, \psi_2 \in \s{B}_{s(x)}(\hat\psi(\bar\theta; x))$,
\[|v(\theta_1, \psi_1; x) - v(\theta_2, \psi_2; x) | \leq b_1(x) (\|\theta_1 - \theta_2\| + \| \psi_1 - \psi_2\|).\]
\item For all $\theta_1, \theta_2 \in \s{B}_r(\bar\theta)$, $\|\hat\psi(\theta_1; x) - \hat\psi(\theta_2; x)\| \leq b_2(x) \|\theta_1 - \theta_2\|.$
\item $b_1$ and $b_1 b_2 \in L_2(P_0)$.
\end{enumerate}
\end{description}

With these conditions we prove Theorem 2.


\begin{proof}[Proof of Theorem 2]
We will use \citet{van2000asymptotic} Theorem 5.23. We need to validate the following conditions to apply the result: (1) $m(\theta; x)$ is measurable as a function of $x$ for all $\theta \in \Theta$; (2) $m(\theta; x)$ is differentiable at $\bar\theta$ for $P_0$-a.e.\ $x$; (3) there exists a measurable function $b \in L_2(P_0)$ and an $r > 0$ such that for all $\theta_1, \theta_2 \in \s{B}_r(\bar\theta)$, $|m(\theta_1; x) - m(\theta_2; x)| \leq b(x) \| \theta_1 - \theta_2\|$; (4) the function $M_0(\theta) = E_{P_0}[ m(\theta; X)]$ is maximized at $\theta = \bar\theta$ and admits a second-order Taylor expansion at $\bar\theta$; and (5) $\tfrac{1}{n} \sum_i m(\hat\theta_n; X_i) \geq \sup_{\theta \in \Theta}  \tfrac{1}{n} \sum_i m(\hat\theta; X_i) - o_P(1)$. We will demonstrate that these conditions follow from conditions (B1)-(B5).

For condition (1), measurability of $x \mapsto m(\theta; x)$ is guaranteed by the measurability of $\hat\psi$ and $v$ plus the fact that compositions of measurable functions are measurable.


Condition (2) is implied by conditions (B2) and (B3) together with the multivariate chain rule. We have $(D_{\theta} m)(\bar\theta; x)= (D_{\theta} v)(\bar\theta, \hat\psi(\bar\theta; x); x) + (D_{\theta} \hat\psi)(\bar\theta; x)^T (D_{\psi}v)(\bar\theta, \hat\psi(\bar\theta; x); x).$ Since $\psi \mapsto v(\bar\theta, \psi; x)$ is maximized at $\hat\psi(\bar\theta; x)$, which is in the interior of $\Psi$, and $v$ is differentiable in $\psi$ at $\theta=\bar\theta$ and $\psi=\hat\psi(\bar\theta; x)$ for $P_0$-a.e.\ $x$, $(D_{\psi}v)(\bar\theta, \hat\psi(\bar\theta; x); x) = 0$ a.s.\ $P_0$. Therefore, $(D_{\theta} m)(\bar\theta; x)=(D_{\theta} v)(\bar\theta, \hat\psi(\bar\theta; x); x)$.

For condition (3), we use (B4). Let $\theta_1, \theta_2 \in \s{B}_r(\bar\theta)$. Then by part (a) of (B4), for each $x$, $\hat\psi(\theta_1; x), \hat\psi(\theta_2; x) \in \s{B}_{s(x)}(\hat\psi(\bar\theta; x))$. Hence, by parts (b) and (c),
\begin{align*}
|m(\theta_1; x) - m(\theta_2; x) | &= | v(\theta_1, \hat\psi(\theta_1; x); x) - v(\theta_2, \hat\psi(\theta_2; x); x) | \leq b_1(x) \left( \|\theta_1 - \theta_2\| + \| \hat\psi(\theta_1; x) - \hat\psi(\theta_2; x) \|\right)\\
&\leq b_1(x) \left( \|\theta_1 - \theta_2\| + b_2(x) \|\theta_1 - \theta_2\|\right)= b_1(x)(1 + b_2(x)) \|\theta_1 - \theta_2\|.
\end{align*}
Since by assumption $b_1, b_1 b_2 \in L_2(P_0)$, condition (3) is satisfied with $b = b_1(1 +b_2)$.

By assumption, $\bar\theta$ is a point of maximum of $M_0(\theta) = E_{P_0}[m(\theta; x)]$. Conditions (B2) and (B3) imply that $m$ is twice continuously differentiable in a neighborhood of $\bar\theta$. Furthermore, we can derive the form of $D_{\theta}^2 m(\theta; x)$ as follows:
\begin{align}
D_{\theta}^2 m(\theta; x) &= D_{\theta} (D_{\theta} m(\theta; x)) =D_{\theta} (D_{\theta} v)(\theta, \hat\psi(\theta; x); x) \\
&=  (D_{\theta}^2 v)(\theta, \hat\psi(\theta; x); x) + (D_{\theta\psi}^2 v)(\theta, \hat\psi(\theta; x); x)  (D_{\theta}\hat\psi)(\theta; x)\label{partial_second_deriv}
\end{align}
By conditions (B1) and (B2), $\hat\psi(\theta; x)$ satisfies $(D_{\psi} v)(\theta, \hat\psi(\theta; x); x) = 0$. Differentiating with respect to $\theta$ gives
\begin{equation} 0 = (D_{\theta\psi}^2 v( \theta, \hat\psi(\theta; x); x) + (D_{\theta}\hat\psi)(\bar\theta; x) (D_{\psi\psi}^2 v)(\theta, \hat\psi(\theta; x); x) \end{equation}
Solving for $(D_{\theta}\hat\psi)(\theta; x)$ and substituting this back in to \eqref{partial_second_deriv} gives
\begin{equation} D_{\theta}^2 m(\theta; x) = \left( D_{\theta\theta}^2 v- (D_{\theta\psi}^2 v) (D_{\psi\psi}^2 v)^{-1} (D_{\theta\psi}^2 v)^T\right)(\theta, \hat\psi(\theta; x); x) .\label{eq:second_deriv}\end{equation}
Therefore, by condition (B3), $|D_{\theta}^2 m(\theta; x)| \leq \kappa(x)$ for all $\theta$ in a neighborhood of $\bar\theta$ and $P_0$-a.e.\ $x$, which implies by the dominated convergence theorem that $M_0$ is twice continuously differentiable in a neighborhood of of $\bar\theta$ with $D_{\theta}^2 M_0(\bar\theta) = E_{P_0}[ D_{\theta}^2 m(\bar\theta; X)]$. Hence, $M_0$ possesses a second-order Taylor expansion at $\bar\theta$, thus satisfying condition (4). 


Finally, condition (5) is satisfied since $\hat\theta_n$ maximizes $ \tfrac{1}{n} \sum_{i=1}^n m(\hat\theta; X_i)$ by definition.

We have now verified the conditions of \citet{van2000asymptotic} Theorem 5.23. Therefore, we can conclude that $\sqrt{n} (\hat\theta_n - \bar\theta)$ converges in distribution to $N_d(0, V(\bar\theta))$, where $V(\theta) = A(\theta)^{-1} B(\theta) A(\theta)^{-1}$ for $A(\theta) =  E_{P_0}[ D_{\theta}^2 m(\theta; X)]$ and $B(\theta) = E_{P_0}\left[ (D_{\theta} m(\theta; X)) (D_{\theta} m(\theta; X))^T\right]$. This establishes the claim.

\end{proof}

\clearpage

\section{Illustrations of the general theory}

\subsection{Consistent and efficient variational estimation}

The first model we study is as follows. $\mathbf{X}_i  = (X_{i1}, \dotsc, X_{ip})$, where conditional on the latent variable $Z_i$, each $X_{ij} \mid Z_i \sim \n{Exp}(Z_i)$, and $Z_i \sim \n{Exp}(\beta)$. Then, with $\theta = \beta$, $p_{\theta}(x, z) = \beta z^p e^{-\left(\beta + \sum_{j=1}^p x_j\right) z}.$ The marginal density of $X_{i1}, \dotsc X_{ip}$ is 
\[ \int_0^{\infty} \beta z^p e^{-\left(\beta + \sum_{j=1}^p x_j\right) z} \, dz = \beta \left(\beta+ \sum_{j=1}^d x_j\right)^{-(p+1)}\int_0^{\infty} u^p e^{-u} \, du = \Gamma(p+1)\beta \left(\beta + \sum_{j=1}^p x_j\right)^{-(p+1)}. \]
The conditional distribution of $Z$ given $X_1, \dotsc, X_d$ is then seen to be Gamma$\left(p+1, \beta+ \sum_{j=1}^p x_j\right)$.

We first validate the derivation of the variational criterion function and profile criterion function stated in the main text. Recall that we take as our variational distribution $q(z; \mu, \sigma) = N(\log z; \mu, \sigma)$. Then
\begin{align*}
 v(\theta, \psi; x) &= E_{\psi} \left[\log\beta + p \log Z - \left(\beta + \sum_{j=1}^p x_j\right) Z  - \log q(Z; \mu, \sigma)\right] \\
 & \propto \log\beta + (p+1)\mu - \left(\beta + \sum_{j=1}^p x_j\right) e^{\mu + \sigma^2 / 2}+ \log \sigma.
 \end{align*}
The function $(p+1)\mu$ is concave in $\mu$, and $\log\sigma$ is strictly concave in $\sigma$. The function $\mu +\sigma^2 / 2$ is convex is $\mu$ and $\sigma$, so $- ce^{\mu + \sigma^2 / 2}$ is strictly concave in $\mu$ and $\sigma$ for any $c > 0$. This shows that $v$ is strictly concave in $\mu, \sigma$ for fixed $\theta$ and $x$. 

The derivative of $v$ with respect to $\mu$ is $(p+1)- \left(\beta + \sum_{j=1}^p x_j\right) e^{\mu + \sigma^2 / 2}$, and the derivative with respect to $\sigma$ is $-\sigma\left(\beta + \sum_{j=1}^p x_j\right) e^{\mu + \sigma^2 / 2} + \sigma^{-1}$. Setting these derivatives to zero and solving the simple resulting system of equations gives $\hat\mu(\theta; x) = \log \frac{p+1}{\beta + \sum_{j=1}^p x_j} - (p+1)^{-1}/2$ and $\hat\sigma(\theta; x) = (p+1)^{-1/2}$. Plugging these expressions in to $v$ gives
\[m(\theta; x) = v(\theta, \hat\psi(\theta; x); x) \propto \log\beta - (p+1) \log \left(\beta + \sum_{j=1}^p x_j \right). \]

Next, we validate the claim that $E[\sup_{\theta} m(\theta; X) ] < \infty$. Differentiating $m$ with respect to $\beta$, we see that the only critical point of $m$ occurs at $\beta = \frac{1}{p} \sum_{j=1}^p x_j$. The second derivative of $m$ with respect to $\beta$ is $-\beta^{-2} + (p+1) \left(\beta + \sum_{j=1}^p x_j \right)^{-2},$ which is negative for $\beta = \frac{1}{p} \sum_{j=1}^p x_j$. Therefore, this critical point is a local maxima, and since it is the only critical point, it is the global maxima of $m(\theta; x)$. Hence, $\sup_{\theta} m(\theta; x) = p \log p - (p+1)\log(p+1) - p\log\left( \sum_{j=1}^p x_j\right)$. Then, using the marginal distribution of $\mathbf{X}$, 
\[ E[\sup_{\theta} m(\theta; x)] \propto -p\Gamma(p+1)\beta \int_0^{\infty} \left(\beta +s\right)^{-(p+1)} \log(s) \, ds < \infty.\]

For the asymptotic variance, we have
\begin{align*}
A(\theta_0) &= E_{P_0} \left[ D_{\theta}^2 m(\theta_0; \mathbf{X})\right] = -\frac{1}{\beta_0^2} + (p+1)E_{P_0} \left[ \left( \beta_0 + \sum_{j=1}^p X_j\right)^{-2}\right] \\
&=  -\frac{1}{\beta_0^2} +\beta_0 (p+1)! \int_0^{\infty} \cdots \int_0^{\infty}  \left( \beta_0 + \sum_{j=1}^p x_j\right)^{-(p+3)} \, dx_1\cdots dx_p \\
&= -\frac{1}{\beta_0^2}  + \frac{\beta_0 (p+1)!}{\beta_0^3 (p+2)(p+1) \cdots 3} = -\frac{p}{p+2}  \beta_0^{-2}.
\end{align*}

\subsection{Inconsistent variational estimation}

We now turn to the second example model that we study in the main text. The setup is identical to the first example, but now $Z \sim \n{Gamma}(\alpha, \beta)$. We have 
\[p_{\theta}(x, z) = \frac{\beta^{\alpha}}{\Gamma(\alpha)} z^{\alpha +p - 1} e^{-\left(\beta + \sum_{j=1}^p x_j\right) z}.\]
 The marginal density of $X_1, \dotsc X_p$ is then
\[ \int_0^{\infty} \frac{\beta^{\alpha}}{\Gamma(\alpha)} z^{ \alpha +p- 1} e^{-\left(\beta + \sum_{j=1}^p x_j\right) z}\, dz  = \frac{\Gamma(p+\alpha) \beta^{\alpha}}{\Gamma(\alpha) \left(\beta + \sum_{j=1}^p x_j\right)^{ \alpha + p} }. \]
The conditional distribution of $Z$ given $X_1, \dotsc, X_p$ is Gamma$\left(\alpha + p, \beta + \sum_{j=1}^p X_j\right)$.

As before, we begin by deriving the variational criterion and profile criterion functions.
\begin{align*}
 v(\theta, \psi; x) &= E_{\psi} \left[\alpha \log\beta - \log\Gamma(\alpha) + (\alpha +p- 1) \log Z - \left(\beta + \sum_{j=1}^p x_j\right) Z  -\log q(Z; \mu, \sigma)\right] \\
 & \propto \alpha\log\beta -\log\Gamma(\alpha)+ (\alpha + p)\mu - \left(\beta + \sum_{j=1}^p x_j\right) e^{\mu + \sigma^2 / 2}+ \log \sigma.
 \end{align*}
 Optimizing with respect to $\sigma$ gives $\hat\sigma^2(\theta; x) = (\alpha + p)^{-1}$. Optimizing with respect to $\mu$ gives $\hat\mu(\theta; x) = \log \frac{\alpha + p}{\beta + \sum_{j=1}^p x_j} - (\alpha + p)^{-1}/2$. Plugging these expressions back in to $v$ gives
 \begin{align*}
  m(\theta; x) &\propto \alpha\log\beta -\log\Gamma(\alpha)-  (\alpha + p)\log\left( \beta + \sum_{j=1}^p x_j \right)- (\alpha + p) + (\alpha + p) \log( \alpha + p) -  \tfrac{1}{2} \log (\alpha + p)\\
  &= \log p_{\theta}(x) - \log \Gamma(p + \alpha)- (\alpha + p) + (\alpha + p) \log( \alpha + p) -  \tfrac{1}{2} \log (\alpha + p).
  \end{align*}
As before, $m$ is smooth in $\theta$, which verifies (A1).  Differentiating $m$ with respect to $\beta$ and solving, we find that for each $\alpha$ and $x$, all critical values of $m$ occurs along $\beta = \alpha \bar{x}$ for $\bar{x} = \sum_{j=1}^p x_j / p$. We can also see that the second derivative of $m$ with respect to $\beta$ is negative at this critical point. Therefore, $\sup_{\theta} m(\theta; x) = \sup_{\alpha} m((\alpha, \alpha \bar{x}); x)$. Some simplification gives 
\[ m((\alpha, \alpha \bar{x}); x) = \alpha \log\alpha - \log \Gamma(\alpha) - (\alpha + p) - \tfrac{1}{2}\log(\alpha + p)  - p\log\bar{x}.\]
Now, $\sum_{j=1}^p X_j \mid Z \sim \n{Gamma}(p, Z)$, so that $E[ \log(\sum_{j=1}^p X_j ) \mid Z] = \psi(p) - \log (Z)$, and therefore by iterated expectation, $E[ -\log(\sum_{j=1}^p X_j )] = -\psi(p) +\psi(\alpha_0) - \log(\beta_0) < \infty$. Furthermore, a basic inequality for the digamma function says that $\log\alpha - \psi(\alpha)  - 1/(2\alpha) > 0$ for all $\alpha > 0$, which implies that $ \alpha \log\alpha - \log \Gamma(\alpha) - (\alpha + p) - \tfrac{1}{2}\log(\alpha + p)$ is strictly increasing. Combining this with the fact that $\Gamma(\alpha) \geq (x/e)^{x-1}$ for $x \geq 2$ yields that $ \alpha \log\alpha - \log \Gamma(\alpha) - (\alpha + p) - \tfrac{1}{2}\log(\alpha + p)  \leq -p-1$ for all $\alpha$. This establishes (A2). (A3) is satisfied if the parameter space is restricted to any compact containing the truth, or via a standard compactification of the parameter space.

\section{Additional simulation results}

\begin{figure}[b]
\centering
\includegraphics[width=6.5in]{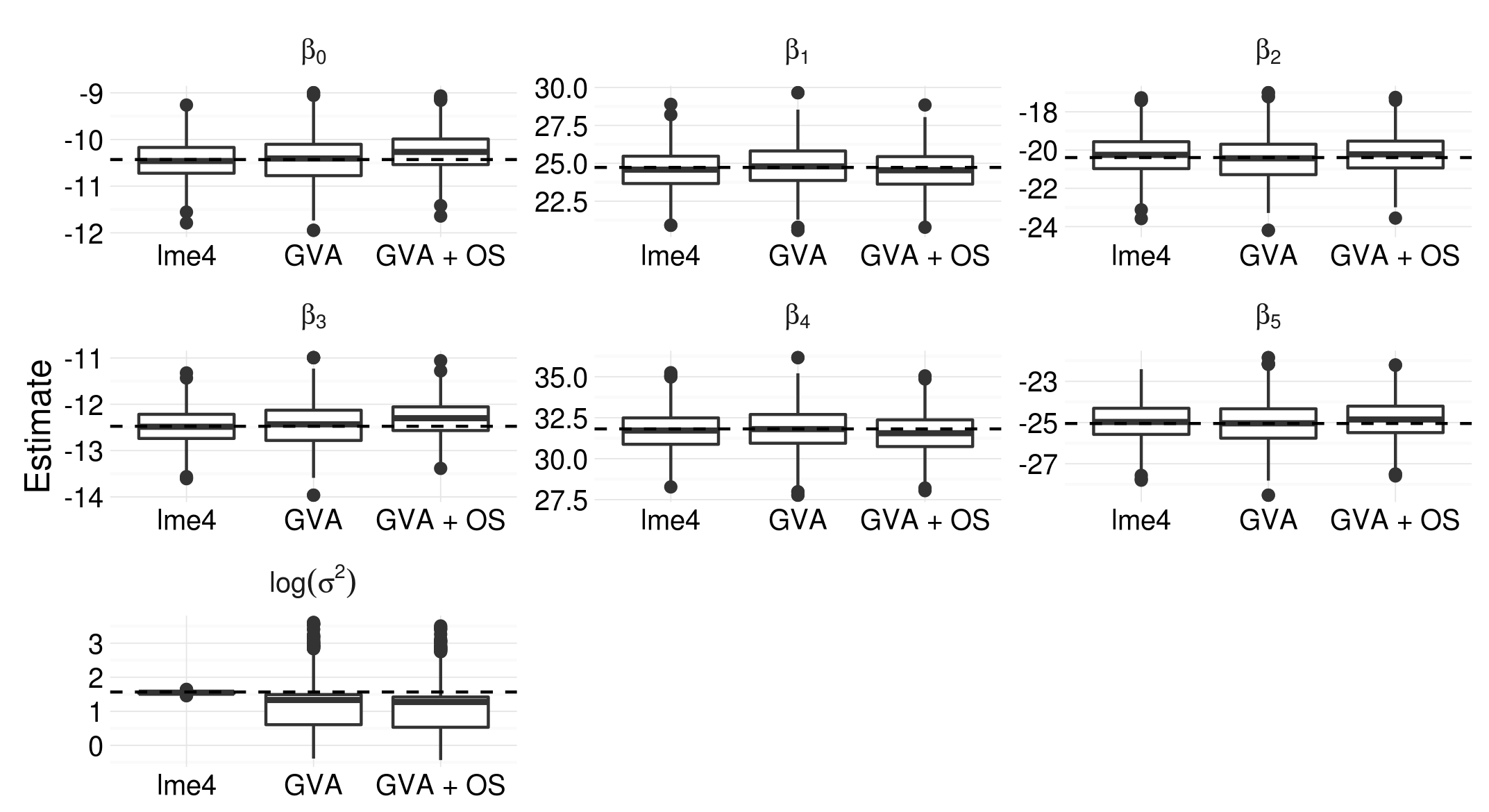}
\caption{Boxplots of parameter estimates from the logistic regression with random intercepts simulation study. ``lme4" corresponds to estimate from the \texttt{lme4} package, ``GVA" stands for Gaussian variational approximation, and ``GVA+OS" refers to the one-step correction.}
\label{fig:rand_int_sim_ests}
\end{figure}

In the main text we tabulated the variances of the three estimators for the random intercepts simulation but did not show the raw estimates. Figure \ref{fig:rand_int_sim_ests} contains box plots of the estimators of each of the seven parameters (the average random effects for females and males $\beta_0$ and $\beta_3$, the fixed linear and quadratic effects of age for females, $\beta_1$ and $\beta_2$, and for males, $\beta_4$ and $\beta_5$, and the log variance of the random effects $\log(\sigma^2)$). 

As discussed in the main text, all three methods appear to be consistent for all elements of $\beta$, but only \texttt{lme4} is consistent for $\log(\sigma^2)$. The raw variational estimates of $\beta$ are slightly less efficient than those of $\texttt{lme4}$ or the one-step correction.

\end{appendix}


\end{document}